\documentclass[twocolumn]{aastex62}

\hypersetup{breaklinks,colorlinks,allcolors=blue}
\usepackage{amsmath}

\usepackage{mathrsfs}

\usepackage[]{units} 

\usepackage{threeparttable}
\usepackage{multirow}
\usepackage[title]{appendix} 

\usepackage{soul}

\usepackage{ulem}

\usepackage{graphicx}				

\def\msun{{\rm\,M_\odot}}

\defcitealias{2017ApJ...834...23S}{Paper I}
\defcitealias{2018ApJ...854...75S}{Paper II}

\begin{document}
\title{Following the Cosmic Evolution of Pristine Gas III:  The Observational Consequences of  the Unknown Properties of Population III Stars }
\shorttitle{Consequences of the Unknown Properties of Pop III Stars}
\author{Richard Sarmento}
\author{Evan Scannapieco}
\affil{School of Earth and Space Exploration, \\
Arizona State University, \\
P.O. Box 871404, Tempe, AZ, \\
85287-1404}
\author{Benoit C\^ot\'e}
\affil{National Superconducting Cyclotron Laboratory, \\
Michigan State University, \\
MI, 48823, USA}
\affil{Konkoly Observatory, \\
Research Centre for Astronomy and Earth Sciences, \\
Hungarian Academy of Sciences, \\
Konkoly Thege Miklos ut 15-17, \\
H-1121 Budapest, Hungary
}
\affil{Joint Institute for Nuclear Astrophysics -- Center for the Evolution of the Elements}
\shortauthors{Sarmento et al.}


\begin{abstract}

We study the observational consequences of several unknown properties of Population III (Pop III) stars using large-scale cosmological simulations that include a subgrid model to track the unresolved mixing of pollutants. Varying the value of the critical metallicity that marks the boundary between Pop III and Population II (Pop II) star formation across 2 dex has a negligible effect on the fraction of Pop III stars formed and the subsequent fraction of Pop III flux from high-redshift galaxies.  However, adopting a log normal initial mass function (IMF) for Pop III stars, in place of a baseline Salpeter IMF, results in a Pop III star formation rate density (SFRD) that is 1/4 of the baseline rate. The flux from high-redshift galaxies modeled with this IMF is highly bimodal, resulting in a tiny fraction of $z \leq 8$ galaxies with more than 75\% of their flux coming from Pop III stars.  However, at $z=9$, right before reionization in our simulations, $\approx$ 20\% of galaxies are Pop III-bright with $m_{\rm UV} \le 31.4$ mag and at least 75\% of their flux generated by Pop III stars .  Additionally, the log normal Pop III IMF results in a population of carbon enhanced, metal poor stars in reasonable agreement with MW halo observations. Our analysis supports the conclusion that the Pop III IMF was dominated by stars in the 20-120$\msun$ range that generate SN with carbon-enhanced ejecta.

\end{abstract}




\section{Introduction}

Many of the  characteristics of the first stars are uncertain \citep{1999ApJ...527L...5B, abel2000formation, 2010AIPC.1294...17N},  but it is likely that they began to form at $z\approx 20$ in minihalos with total masses $\sim 10^{6} \msun$ \citep{abel2002formation, Bromm2002}. Since H$_{2}$ was the only available coolant in the primordial gas, fragmentation was likely mostly suppressed giving rise to an initial mass function (IMF) that was likely biased toward massive stars \citep{2006ApJ...641....1T, 2007ApJ...661...10B,2010MNRAS.401L...5S,2014ApJ...792...32S}.

The shape and characteristic mass of the  Population III (Pop III)  IMF are still unknown although estimates of the likely maximum mass of such stars have been decreasing in recent years \citep{Whalen2012a, 2013ApJ...773..185S, 2014ApJ...792...32S, 2018ApJ...857...46I}. Early modeling efforts estimated the mass of the first Pop III stars at 100-500 $\msun$ \citep{1999ApJ...527L...5B, 2001MNRAS.328..969B, abel2000formation, abel2002formation}. More recent estimates point to an IMF in the range of 30-300 $\msun$ \citep{o2007population}.
However, there is now evidence from numerical simulations of primordial proto-stellar clouds that the very first Pop III stars did not necessarily form solely as single stars or binaries. Instead, the proto-stellar accretion disk may have fragmented resulting in the birth of several Pop III stars from a primordial natal cloud  with masses of order 10$\msun$ \citep{2009Sci...325..601T, 2010MNRAS.403...45S, 2010MNRAS.404.1425J, 2011ApJ...727..110C, 2013MNRAS.433.1094S, stacy2014first}.

Although there have been several efforts to find Pop III stars in and around the Galaxy as well as in early galaxies at high redshift, observational efforts have yet to discover this first generation of stars \citep{2008ApJ...680..100N, 2012ApJ...761...85K, 2013A&A...556A..68C, 2015ApJ...808..139S}.  Given the lack of observations of Pop III stars, it has fallen to theory and simulation to derive some of their characteristics.  While theoretical studies have become more sophisticated \citep{2003ApJ...586....1M, 2003ApJ...589...35S,o2007population,  2007MNRAS.382..945T,2010AIPC.1294...17N, wise2011birth, Johnson2013a, Hartwig2014, 2014MNRAS.440.2498P, 2016ApJ...823..140X, Jaacks2018}, they  have to make simplifying assumptions and choices for several of the relevant parameters surrounding star formation and supernova (SN) feedback given the scales required to model a statistically significant volume in the early universe.

The most important of these are likely to be: 1) the critical metallicity that marks the boundary between Pop III and Pop II star formation and 2) the efficiency of Pop III SNe feedback that determines how metals from the first stars are disseminated.
The critical metallicity is believed to be between $10^{-6}$ and $10^{-3} Z_{\odot}$ depending on how efficiently dust can act to radiatively cool the gas \citep{2011ApJ...729L...3D, 2012MNRAS.419.1566S, 2013ApJ...766..103D}. Its value has direct implications for Pop III star formation since it dictates which parcels of gas form Pop III stars. The energy generated by Pop III SNe \citep{2005ApJ...630..675K, 2014ApJ...797...97S}, on the other hand, is tightly coupled to the Pop III IMF, which determines the fraction of stars that generate SNe and the energies with which they eject metals.

Together these unknown properties have important implications for the two main methods used to constrain Pop III star formation: direct searches and 
analyses of extremely metal poor (EMP) stars. While direct measurements of individual Pop III stars at high redshift are challenging \citep{windhorst2018}, dwarf galaxies containing these stars are expected to display several unique signatures. Pop III stars are predicted to be very blue, not only because they are likely to be massive but also because they have lower opacities and higher core temperatures \citep{1983ApJ...272..249C, 1983A&A...119...54E}.
This leads to stronger UV emission, inducing strong nebular Lyman-$\alpha$ emission \citep{2001MNRAS.328..969B, 2002A&A...382...28S}, as well as nebular HeII 1640 Angstrom emission, which is one of the most promising Pop III tracers \citep{2000ApJ...528L..65T, 2001ApJ...550L...1T, 2003RMxAC..17..230P}. 

A complementary approach to constraining the properties of the first stars is to analyze their fossil remnants. While Pop III stars themselves are not expected to last until the present day, the elements they ejected should remain preserved in low-metallicity, second-generation stars, unmixed with material from later populations. These second generation stars are rare overall, but are expected to make up a substantial fraction of  metal-poor stars
\citep{2002ARA&A..40..487F, 2005Natur.434..871F, 2015ARA&A..53..631F}.
Of particular interest are  CEMP-no\footnote{CEMP-no: [Fe/H] $< -1.0$, [C/Fe] $>+0.7$, and [Ba/Fe] $< 0.0$} stars, Carbon Enhanced Metal Poor (CEMP) stars that do not display an enhancement of  \textit{r} or \textit{s}-process elements \citep{2005ARA&A..43..531B}.
Such stars are likely polluted by a single, or at most a few Pop III SNe.  Their unusual abundances argue for a Pop III IMF with a characteristic mass of order 10's of solar masses \citep{2014Natur.506..463K, 2015MNRAS.447.3892H, 2017MNRAS.465..926D, 2018ApJ...857...46I}.  If this is true, the overall Pop III rate of Type II SNe would be significantly higher than that of stars today  \citep{2002RvMP...74.1015W, 2002ApJ...567..532H,2013ApJ...768...95W}.

In \cite{2017ApJ...834...23S}, hereafter Paper I, we attempted to identify a  characteristic Pop III  mass by comparing the SN nucleosynthetic yields of Pop III stars to the chemical abundances measured in a subset of carbon enhanced, metal poor Milky Way (MW) halo stars.   In \cite{2018ApJ...854...75S}, hereafter Paper II,  we characterized the high redshift luminosity function (LF) \citep{2015ApJ...807L..12O, 0004-637X-813-1-21}, making predictions for the James Webb Space Telescope (JWST). Both these studies used a value near the middle of the range for the critical metallicity, $Z_{\rm crit} = 10^{-5}$, along with typical SN rates and energies for stellar masses drawn from a \cite{salpeter1955luminosity} IMF\footnote{While SN rates were based on a Salpeter IMF, the flux from the surviving stars was modeled, in post-processing, using a log-normal IMF. This model mismatch is addressed in this work.}. All Pop III SN yields were assumed to be from a $60 \msun$ Pop III star \citep{hegerwebsite}. 

In this work, we vary several of the parameters discussed above, evaluate the implications against our previous results, and compare them to observations of CEMP-no stars. In particular, we investigate the effects of the value of the critical metallicity on the transition from Pop III to Pop II star formation. We implement a log-normal IMF, within our simulation, for Pop III stars that significantly changes the amount of SN feedback and the fraction of surviving Pop III stars after $\approx$~10~Myr. We also examine the effects of different SN mass loading on subsequent Pop III star formation.  

This paper is structured as follows.  In Section \ref{sec:meth} we describe our methods, including a brief review of our implementation of the subgrid model tracing the fraction of pristine gas, some of the relevant physics modeled, and the setup for the simulations used in this parameter study. In Section \ref{SEC:res} we discuss our results and the effects of the various parameters on the high-redshift luminosity function (LF), the fraction of Pop III flux emitted by early galaxies, and the subsequent effect on the chemical composition of Pop II stars. Finally, conclusions are discussed in Section \ref{SEC:con}. 

\section{Methods\label{sec:meth}}
In this section, we review the simulation methods and  parameter choices for the runs discussed in this work. The physics, cosmology and the implementation of the pristine gas fraction scalar, as well as the primordial metallicity scalar, are described in more detail in \citetalias{2017ApJ...834...23S}. 

\subsection{Simulation Setup\label{sec:simsetup}}

To facilitate comparisons with \citetalias{2017ApJ...834...23S} and \citetalias{2018ApJ...854...75S}, we adopted the cosmological parameters used therein: $\Omega_{\rm M} = 0.267$, $\Omega_{\Lambda} = 0.733$, $\Omega_{\rm b} = 0.0449$, $h = 0.71$, $\sigma_8 = 0.801$, and $n = 0.96,$ based on \cite{2011ApJS..192...18K}. These parameters have their usual cosmological definitions. 

We again made use of \textsc{ramses} \citep{2002A&A...385..337T}, an adaptive mesh refinement (AMR) cosmological simulation code, to model a 12 Mpc $h^{-1}$ on-a-side volume generated from Multi-Scale Initial Conditions (MUSIC) \citep{2013ascl.soft11011H}. We evolved this volume to $z=7$ covering approximately the first 780 Myr of cosmic history, given our cosmology, for each of the new simulations analyzed. 

The following parameters are common to all of the simulations discussed in this work. The initial grid resolution of 11.7 comoving kpc h$^{-1}$ was based on a starting resolution of $1024^{3}$ cells ($\textit{l}_{\rm min} = 10$). Our choice for the initial resolution resulted in a dark matter (DM) particle mass of $4.47 \times 10^{5}\, \msun\, h^{-1}\, \Omega_{\rm \textsc{dm}}$ or $1.40 \times 10^{5}\, \msun$ for our cosmology. 

We took a quasi-Lagrangian approach to refinement. As cells became overdense, \textsc{ramses} refined them to ensure the mean mass per cell was roughly constant across the simulation. We allowed for up to eight additional refinement levels ($\textit{l}_{\rm max} = 18$); however the highest refinement level reached by $z=7$ was $\textit{l} =15$ resulting in a best physical resolution of 64.5 pc.
 
We did not model black holes (BHs), since their feedback is likely not important for high-redshift galaxy evolution \citep{2004ApJ...608...62S, 2008MNRAS.391..481S,2012AIPC.1480..325J,2017ApJ...836..216P}, the galaxy mass range explored or for the parameters we are interested in investigating.  The redshift of reionization was also a free parameter, which we set to $z_{\rm re}=8.5$, as reported by the \cite{2016A&A...594A..13P}. Reionization is a discrete event in the simulation.

\subsection{Simulation Physics}

\textsc{Ramses} uses CLOUDY \citep{1999ascl.soft10001F} to model atomic/ionic cooling for $T \gtrsim 10^{4}$ K. We used the cooling rates from \cite{rosen1995global} for temperatures below this value. We set the radiative cooling floor to 100 K, but adiabatic cooling can lower the gas temperature below this value. The UV background was derived from \cite{1996ApJ...461...20H}.

As discussed in \citetalias{2017ApJ...834...23S}, we also modeled molecular cooling in the pristine gas \citep{johnson2006cooling, 2008arXiv0809.2786P, hirano2013radiative}. Our analytic model is based on work by \cite{martin1996master} and provides a radiative cooling rate, $\Lambda_r/n_{H_2}$, for the gas densities encountered in the simulations. We assumed a primordial H$_{2}$ fraction of 10$^{-6}$ \citep{2005MNRAS.363..393R} and an optically thin simulation box. As a result, all of our H$_{2}$ is destroyed shortly after the first stars are formed since they produced more Lyman-Werner photons than there were H$_{2}$ molecules in our simulation volume. We did not model H$_{2}$ formation.

The simulations spawned star particles (SPs) in regions of gas according to a Schmidt law \citep{1959ApJ...129..243S} using a star formation rate 
\begin{equation}\label{eqn:sf}
\frac{d\rho_{\star}}{dt} =\epsilon_{\star} \frac{\rho}{t_{\rm ff}},\; \rm {when}\,  \rho > \rho_{\rm th}, 
\end{equation}
where  $t_{\rm ff} = \sqrt{3 \pi /(32 G \rho)}$ is the gas free fall time. Star forming regions also needed to be at least 200 times the mean density of the simulation, as a function of redshift, before SPs were formed. This ensured that SPs were only formed in collapsed objects and not in high-density flows \citep{2006A&A...445....1R,2008A&A...477...79D}.  

For our simulations we set $\rho_{\rm th} = 1.0\, H\, {\rm cm}^{-3}$ and the star forming efficiency to $\epsilon_{\star} = 0.01$, giving results in reasonable agreement with the observed high redshift star formation rate \citep{2016PASA...33...37F, 2014ARA&A..52..415M}. Our SP mass resolution was 
\begin{equation}\label{eqn:spmr}
m_{\star} = \frac{\rho_{\rm th}}{X} \Delta x_{\rm min}^{3} = 8.6 \times 10^{3}\, \msun, 
\end{equation}
where we account for a primordial H fraction of $X=0.76$. The final mass of each SP was drawn from a Poisson process such that it was a multiple of $m_{\star}$.

The range of SP masses generated across the simulations was $8.6\times10^3\, \msun \leq M_{\star} \leq 6.0\times 10^4\, \msun$. When considering our log normal IMF and the minimum SP mass, we sampled 62 stars in the mass range 20-120$\msun$ comprising a total mass of $\approx 3580\msun$ per SP. Additionally, all of our star forming galaxies at $z\le 16$ generated at least 4 SPs/halo. The nonlinear length scale at the end of the simulations, $z=7$, was $47$ comoving kpc h$^{-1}$. This scale corresponds to a mass of $3.2\times 10^{7} M_\odot$ h$^{-1}$.


Each SP formed prompts a SN event after 10 Myr. This SN represents the stars with masses greater than approximately 16$\msun$ in the IMF. We note that all of our SN feedback occurred after 10 Myr even though many of the very high mass stars would have had much shorter lives. Further, each SN injected $E_{\rm SN}=10^{51}$ ergs/10 $\msun$ in the form of kinetic energy of the gas.

Radiation pressure from Pop III and other massive stars can disrupt -- and also trigger -- star formation \citep{whalen2004radiation,  deharveng2010gallery, wise2011birth, 2012A&A...546A..33T}, affecting the star formation rate in high-redshift galaxies, we did not include its effects in this work.

\subsection{The Pristine Fraction, Corrected Metallicity and the Primordial Metallicity}\label{sec:PF}
For these simulations, we made use of the modifications to \textsc{ramses} described in \citetalias{2017ApJ...834...23S} to track two new metallicity-related scalars. These allowed us to more accurately model the fraction of Pop III stars created in high-redshift galaxies and to follow the unique elemental yields generated by Pop III SN.  The \textit{pristine gas mass fraction}, $P$, and the \textit{primordial metallicity}, $Z_{\rm P}$, are described below. Additionally, we discuss how we used the pristine fraction to improve the accuracy of the metallicity of the polluted fraction of gas and SPs.

\subsubsection{The Pristine Fraction}
The pristine gas fraction, $P$, was used to track the mass fraction of gas with $Z < Z_{\rm crit}$ in each simulation cell. The scalar evolves from $P=1$, indicating that 100\% of the gas in the cell is metal-free, to $P=0$, indicating that all of the gas in the cell has been polluted above $Z_{\rm crit}$.  $P_{\star}$ records the value of $P$ in SPs at the time they were spawned and identifies the mass fraction of Pop III stars with $Z_{\star} < Z_{\rm crit}$ for each SP. 

As discussed in detail in \cite{2013ApJ...775..111P}, and first implemented in a cosmological simulation in \citetalias{2017ApJ...834...23S}, the following equation was used to describe the evolution of the pristine gas fraction:
\begin{align}\label{eq:selfConv}
\frac{d P}{d t} = - \frac{n}{\tau_{\rm con}}  P\left(1-P ^{1/n}\right). 
\end{align}

The change in the pristine gas fraction at each time-step was therefore a function of $n$, a measure of the locality of mixing, and a timescale, $\tau_{\rm con}$,  that is the inverse of the turbulent stretching rate \citep{2010ApJ...721.1765P, 2012JFM...700..459P, 2013ApJ...775..111P}. These parameters are, in turn, functions of the turbulent Mach number, $M$, and the average metallicity of the cell relative to the critical metallicity, $\overline Z /Z_{\rm crit}$ \citepalias{2017ApJ...834...23S}. By knowing $P$ at the time of star formation we subsequently modeled the mass fraction of Pop~III stars for each SP formed as $M_{\star, \textsc{\tiny III}} = M_{\star}\times P_{\star}$. 

\subsubsection{The Corrected Metallicity}\label{sec:cm}
Each SP in the simulation recorded both the host cell's average metallicity, $\overline Z \rightarrow \overline Z_{\star}$, and the pristine gas fraction, $P\rightarrow P_{\star}$ at the time it was created. This information was combined to model the metallicity of the polluted fraction of gas or stars. When metals were well mixed throughout the cell's volume $P=0$ and the scalar $\overline Z$ represented the average metallicity of any sub-volume of gas in the cell. However, before mixing was completed, the polluted fraction, $f_{\rm pol} \equiv 1-P$, described the fraction of gas that was actually polluted with metals. Therefore, a better estimate of the metallicity of the polluted fraction of gas was the \textit{corrected metallicity},
\begin{equation}\label{eq:zcorr}
\begin{aligned}
Z = \frac{\overline Z} {f_{\rm pol}}.
\end{aligned}
\end{equation}

As expected, when $f_{\rm pol} = 1$ the corrected metallicity was the average metallicity. However, when $f_{\rm pol} < 1$ only a fraction of the cell was polluted and the metals were concentrated in a volume smaller than the cell and hence $Z > \overline{Z}$ for that fraction of gas. When referring to the corrected metallicity of the polluted fraction of gas or SPs, we use $Z$ or $Z_{\star}$.

As discussed in \citetalias{2017ApJ...834...23S}, the corrected metallicity was precise only when all of the metals in the cell are contained in the polluted fraction. However, it is possible for the gas in a cell to have been uniformly polluted to $0 < \overline Z < Z_{\rm crit}$ via advection or mixing of dilute metals from a previous SN. Such cells are still capable of creating Pop III stars. When material from a new SN entered such a cell, the polluted fraction, $f_{\rm pol}$, did not represent all of the metals in the cell.

Equation (\ref{eq:zcorr}) accurately models the metallicity of the polluted fraction of gas when the unpolluted fraction has $\overline{Z}=0$. This equation is therefore the upper bound to the metallicity of the polluted fraction. A lower bound to the corrected metallicity, for the situation described above, is reached when the maximum fraction of metals is already well mixed in a cell but the cell is just below $Z_{\rm crit}$. When new pollutants entered such a cell it resulted in the metallicity of the polluted fraction
\begin{equation}\label{eq:lowerlim}
\begin{aligned}
Z = \frac{\overline Z - Z_{\rm crit}}{f_{\rm pol}},
\end{aligned}
\end{equation}
where $Z_{\rm crit} = \{10^{-6},\, 10^{-5},\,  10^{-4}\} Z_{\odot}$ (for runs Z6, fid and Z4, respectively). This equation accounts for metals in the pristine fraction of gas. Note that when we set $Z_{\rm crit}=0$, representing a cell with purely primordial gas in its pristine fraction, equation (\ref{eq:lowerlim}) becomes equation (\ref{eq:zcorr}) as we would expect. Given the small amount of metals that could be mixed in the pristine fraction, we ignore this effect for cells that had a pristine fraction $0 < \overline Z < Z_{\rm crit}$. 

Finally, we did not create polluted stars when $f_{\rm pol} <  10^{-5}$. This limit had no impact on our final results even though imposing it could have theoretically resulted in a tiny fraction of high enriched SPs. However, the maximum metallicity reached for any SP in the simulation was 0.36 Z$_{\odot}$ at z = 7.

\subsubsection{Primordial Metals}
We refer to the metals generated by the first stars as \textit{primordial metals}. Since the elemental abundance patterns for ejecta from massive Pop III SN are likely different from those of lower-mass Pop II stars \citep{2002ApJ...567..532H}, we developed a straightforward method to track these metals in a cosmological simulation. Anytime a SP with a non-zero pristine fraction went SN, we tracked the Pop III SN ejecta with the scalar $\overline Z_{\rm P}$.

This new scalar allowed us to follow primordial metals injected into the gas and to model the final metal content of subsequently formed SPs. Once we applied our correction to the metallicity, the fraction of primordial metals in SPs was computed as $Z_{\rm P, \star}/Z_{\star}$ where $Z_{\star}$ captures the total metallicity of the SP. When mapping these metallicities to the chemical composition of a SP (or fraction thereof for $0 < P < 1$), we used $Z_{\star}-Z_{\rm P, \star}$ to model the mass fraction of `normal' metals, while $Z_{\rm P, \star}$ modeled the mass fraction of primordial metals produced by a representative Pop III SN abundance pattern or integrated Pop III IMF SN yields.

\subsection{Simulations}\label{sec:SR}
We carried out a set of 5 large-scale cosmological simulations to study the effects of varying the critical metallicity, SN mass loading, and  Pop III IMF on the star formation rate, the luminosity of galaxies as well as the fraction of Pop III flux coming from them. Table \ref{tab:sims} identifies the simulations and their corresponding parameters. Note that the fid and fid\_orig runs use the same simulation output: the only difference is the IMF used to model surviving (post-SN) SP mass. All other simulation parameters were consistent across all runs, as described in Section \ref{sec:simsetup} and captured in Table \ref{tab:simsCommon}.

We include comparisons to the simulation in \citetalias{2018ApJ...854...75S}, fid\_orig, that used a log normal IMF to model the spectral energy density (SED) of Pop III stars in post-processing, even though the surviving post-SN SP mass fraction, after 10 Myr, was representative of a Salpeter IMF.

\begin{table}
\caption{\label{tab:sims}Simulation Parameters} 
\begin{tabular}{r|cccccc}
\hline
\hline
Name & $Z_{\rm crit}$ & $f_{w}$\tablenotemark{a} & Pop III IMF & $\alpha$ & $\nicefrac{M_{\rm c}}{\msun}$ & $\sigma$ \\
\tableline
fid & $10^{-5}$ & 10 & Salpeter\tablenotemark{b,d} & 2.35& -- & -- \\
Z4 & $10^{-4}$ & 10 & Salpeter\tablenotemark{b} & 2.35 & -- & -- \\
Z6 & $10^{-6}$ & 10 & Salpeter\tablenotemark{b} & 2.35 & -- & -- \\
P3SN & $10^{-5}$ & 10 & Log normal\tablenotemark{c} & -- & 60 & 1.0 \\
fw1 & $10^{-5}$ & 1 & Salpeter\tablenotemark{b} & 2.35 & -- & -- \\
fid\_orig & $10^{-5}$ & 10 & Log normal\tablenotemark{c,d} & -- & 60 & 1.0 \\
\tableline
\end{tabular}
\tablenotetext{a}{SN mass loading. The amount of gas, expressed in multiples of the mass of SN ejecta, carried along with the SN blast and removed from the host cell.}
\tablenotetext{b}{The Salpeter IMF mass range spans 0.6 to 100 M$_{\odot}$.}
\tablenotetext{c}{The log normal IMF  mass range spans 1 to 500 M$_{\odot}$. }
\tablenotetext{d}{Both fid and fid\_orig are based on the same \textsc{ramses} simulation data. However, for fid the SEDs used to model Pop III flux in post-processing assumed a log normal IMF for SP mass; the fid\_orig results assumed a Salpeter IMF. }
\end{table}

The critical metallicity, $Z_{\rm crit}$, marks the boundary between Pop III and Pop II star formation. This parameter is loosely constrained \citep{2005ApJ...626..627O} and we explored the effects of varying the critical metallicity across 2 orders of magnitude. We defined the critical metallicity of the gas for the fiducial (fid) run as $Z_{\rm crit} = 10^{-5} Z_\odot$, as was done in our earlier works (fid\_orig). The runs Z4 and Z6 were used to determine the effects of setting the critical metallicity 1 dex higher, and lower, respectively.

By default, the \textsc{ramses} model for stellar evolution assumes that a user-specified fraction, $\eta_{\rm SN}$, of the mass of each SP, regardless of the SP's metallicity or pristine fraction, goes SN after 10 Myr. We used $\eta_{\rm SN}= 0.10$ for the fid, Z4, Z6, fw1, and fid\_orig runs corresponding to the massive, short lived stars at the top of a \cite{salpeter1955luminosity} IMF with a mass range of 0.6 to 100$\msun$.  

For this study we implemented a second IMF for Pop III stars within the simulation. The fraction of each SP, in run P3SN, with metallicity below $Z_{\rm crit}$ is $P_{\star}\times M_{\star}$ and was modeled using a log normal Pop III IMF \citep{larson1973simple, 2010A&A...523A..64R,2006ApJ...641....1T}. Our Pop III IMF had a characteristic mass of 60 M$_{\odot}$ and $\sigma = 1$. The mass range spanned 1 to 500 M$_{\odot}$ and corresponds to Model ID ``TE'' in \cite{2010A&A...523A..64R}. Specifically, our IMF was
\begin{equation}\label{eq:halo}
\begin{aligned}
\frac{dN}{d\,\mathrm{ln}M} = \exp\left[-0.902 - \frac{1}{2\sigma^{2}}\, \mathrm{ln} \left(\frac{M}{60 \msun} \right)^{2}\right],
\end{aligned}
\end{equation}
where -0.902 normalizes the PDF.

Our Pop III IMF resulted in 99\% ($\eta_{\rm SN, III}= 0.99$) of the stars going SN within the first 10 Myr \citep{2010ApJ...724..341H} representing a significant increase in feedback energy as well as the amount of metals injected into the ISM as compared to the fiducial case. Note that we used the same Type II SN fractional metal yield and explosion energy for all Pop III SN in this IMF although a subset of stars in the mass range 40-140$\msun$ may collapse directly into BHs while higher mass stars likely generate pair-instability SN  with different explosion energies and yields \citep{2003ApJ...591..288H,2005ApJ...633.1031S}. We leave the exploration of this parameter space to a future work.

Still considering run P3SN, the fraction of Pop II stars, per SP, had mass $(1.0 - P_{\star})\times M_{\star}$. This IMF was modeled using the Salpeter IMF described above with $\eta_{\rm SN}= 0.10$. The Pop II IMF used the standard power-law slope, $\alpha=2.35$.

As mentioned above, all SN in the simulation injected $E_{\rm SN}=10^{51}$ ergs/10 $\msun$ as kinetic energy \citep{2008A&A...477...79D}.  Further, we assumed that 15\% of SN ejecta, by mass, were metals, regardless of the type of SP.

The mass-loading parameter, $f_{w}$, established the amount of gas, in terms of a multiple of the mass of SN ejecta, in the host cell that was carried along with the SN blast. For this work, and our previous studies, we used a default value of $f_{w} = 10$. This material was subsequently distributed equally among all of the cells within the SN blast radius of 300 pc, or a volume of $\approx1.13\times 10^{8}$ pc$^3$. For this study, we explored the effect of reducing the mass loading factor to $f_{w}=1$ via run fw1.

While the fractional difference in the amount of material carried into the cells in the SN blast radius was insignificant, reducing $f_{w}$ from 10 to 1 resulted in high-redshift galaxies retaining more of their gas.  Specifically, for the $f_{w}=10$ fiducial case, each SN removed 1 SP mass of gas, at least $8.6\times10^{3}\msun$, from the host cell since the SN ejecta mass was 10\% of the SP mass. This helped to modulate future star formation by reducing the gas in star forming cells, although, for numerical reasons, the maximum amount of gas carried by any SN was capped at 25\% of the gas remaining in the host cell. For $f_{w} = 1$, 10x more gas was left in the central star forming region of active galaxies as compared to the fiducial case, assuming the 25\% limitation was not routinely reached.

\begin{table}
\caption{\label{tab:simsCommon}Common Simulation Parameters} 
\begin{center}
\begin{tabular}{|r|l|r|l|}
\hline
\hline
Name & value & Name & Value \\
\tableline
$\Delta x_{min}$ & 64.5 pc & $\rho_{th}$ & 1 $H\; cm^{-3}$ \\
$M_{DM}$ &  $1.4 \times 10^{5} \msun$ & $E_{SN}$ & 10$^{51}$ erg / $10\msun$\\
$m_{\star}$ & $8.6 \times 10^{3} \msun$ & $z_{re}$ & 8.5 \\
$r_{bubble}$ & 300 pc & SN Z yield\tablenotemark{a} & 0.15 \\
\tableline
\end{tabular}
\end{center}
\tablenotetext{a}{The mass fraction of metals produced by SNe.}
\end{table}

\subsection{Halo Finding}
AdaptaHOP \citep{2004MNRAS.352..376A} was used to find star forming galaxies in the simulation. Groups of 20 particles were used to compute the local matter density and halos must have been composed of at least 100 DM particles, equivalent to a DM halo mass of $1.4 \times 10^{7}\, \msun$, to have been considered a candidate halo.  Further, only candidates with densities 80 times the average total matter density, as a function of redshift, were stored.

\begin{figure*}[t!]
\begin{center}
\includegraphics[width=1\textwidth]{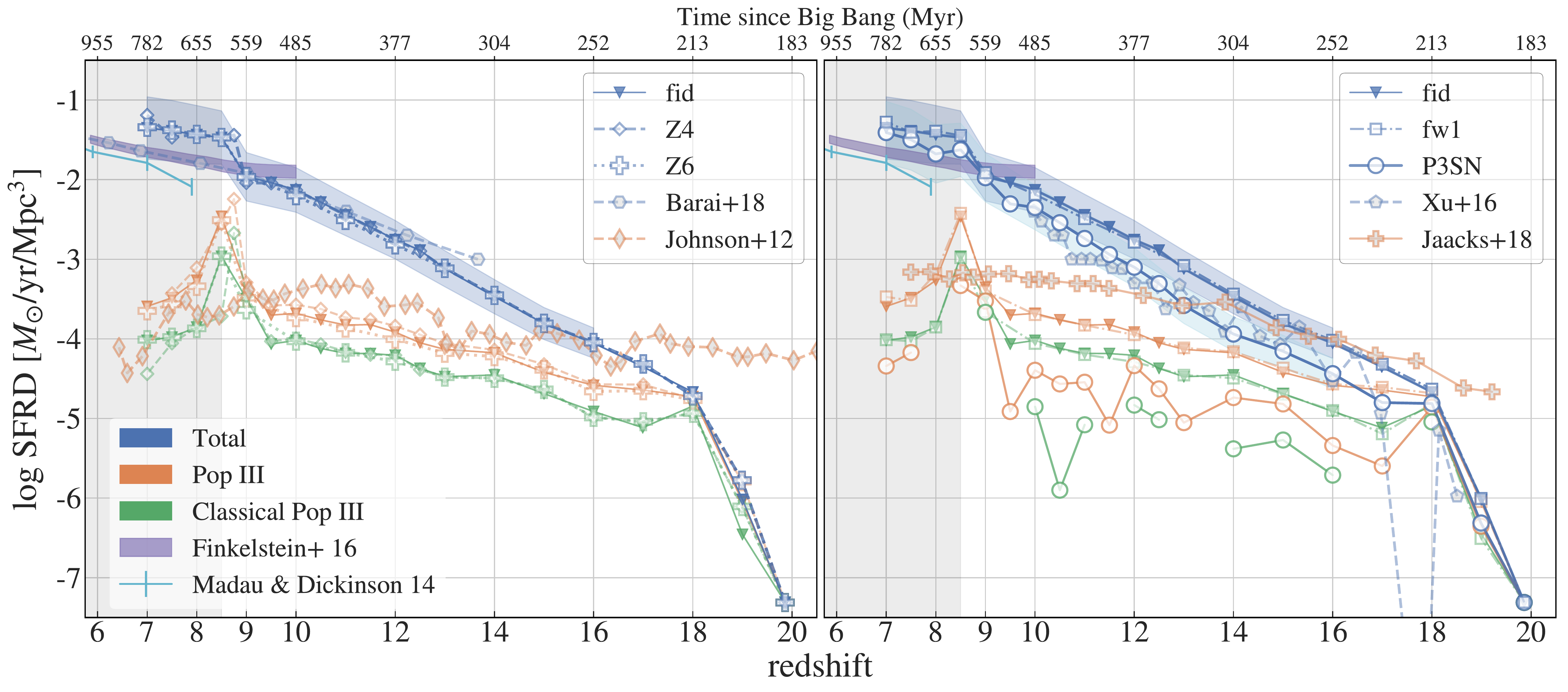}
\caption{SFRDs for our simulations and, for comparison, from \cite{2018arXiv180704768B, 2016ApJ...833...84X} along with a compilation of observations by \cite{2014ARA&A..52..415M}. We also include a LF-derived SFRD by \cite{2016PASA...33...37F}, based on an integration of the reference's LF to $M_{\rm UV}$ = -13 mag. The 1$\sigma$ uncertainties on the fid and P3SN runs (light blue shading) at $z \le 16$ accounts for both Poisson noise and sample variance. We include two Pop III SFRDs for each of our simulations. The ``Pop III'' (orange) rate includes the effects of the subgrid model when determining the fraction of Pop III stellar mass. The ``classical Pop III'' (green) rate only includes Pop III SPs that are formed in cells with $Z < Z_{\rm crit}$. Pop III SFRDs by \cite{Jaacks2018, 2013MNRAS.428.1857J}, are also in orange. The Pop III SFRD show very little variation over 2 orders-of-magnitude of the critical metallicity (runs Z4 \& Z6). The largest effect on the star formation rate occurs when changing the IMF for Pop III stars; run P3SN. The Pop III SFRD for run P3SN is $\approx$0.23 of the fid rate owing to both increased feedback from the Pop III IMF and to the smaller surviving fraction of Pop III stars. The grey-shaded area indicates redshifts post-reionization. If data is missing for a redshift it indicates a SFRD of zero. \label{fig:sfrd3}}
\end{center}
\end{figure*}

Many of the more massive objects found by AdaptaHOP were composed of more than one observationally distinguishable galaxy and these overly bright objects had the potential to bias the bright end of our LFs. To determine the radius of observationally distinguishable galaxies we post-processed them as described in \citetalias{2018ApJ...854...75S}. Our method assumed an HST-like criteria for separability of sources and required at least a 0.1 arcsec separation between objects. The vast majority of these objects were too dim to detect in unlensed JWST fields. However they did provide information at the faint end of our LFs. 

\subsection{Galaxy Spectral Models and Simulated Observations}\label{SEC:gsm}
We computed the UV fluxes, at a rest-frame wavelength of 1500 $\AA$, for our simulated galaxies based on the redshift, ages, metallicities, and masses of their constituent SPs using a set of simple stellar population SED models as was done in \citetalias{2018ApJ...854...75S}. Our original study, fid\_orig, modeled the flux from all Pop III stars with $Z_{\star} < Z_{\rm crit}$ using $Z_{\star}=0$ SEDs using a log normal IMF to interpret SP mass. This was in tension with the Salpeter IMF modeled in the simulation. While we pointed this out in the original work, we correct this mismatch by using SEDs based on a Salpeter IMF for new runs that use 10\% as the SN fraction for SPs. This fraction corresponds to the massive stars at the top of a Salpeter IMF. This includes runs fw1, fid, Z4, and Z6. For run P3SN we use the log normal SED for Pop III SPs that agrees with the 99\% of SP mass that goes SN after 10 Myr.

Each of the model SEDs were redshifted over the range $z$=7-16 and attenuated to account for Lyman forest and continuum absorption \citep{1995ApJ...441...18M}. This process also included a spectral conversion from wavelength to frequency converting the SEDs into the familiar units of flux, normalized to an IMF of stars with 1 solar mass: ${\rm erg/s/Hz/cm}^{2}/\msun$. As in our previous work, we did not consider attenuation due to dust.

All of the SEDs, translated into fluxes at a given redshift, modeled stars of discrete metallicities and discrete ages spanning the age range of SPs in the simulation. We interpolated the data in both metallicity and age linearly, in log-space, to determine  the flux of specific SPs and fractions thereof.

Specifically, the fluxes for all Pop II SPs, for all runs, were based on an SED with a Salpeter IMF normalized to $1 \msun$. The fluxes were derived from \textit{STARBURST 99} SEDs \citep{2014ApJS..212...14L}, supplemented with SEDs by \cite{2010A&A...523A..64R}, for $Z_{\rm crit} \le Z_{\star} \le 5\times 10^{-4}\, Z_{\star}$.

Pop III SPs, with $Z_{\star} < Z_{\rm crit}$, had fluxes that were based on \cite{2010A&A...523A..64R} SEDs for a stellar population with $Z_{\star}=0$. For runs fid, fw1, Z4, and Z6 the SEDs modeled a Salpeter IMF. Run P3SN used a Pop III SED modeled on the log normal IMF described in Table \ref{tab:sims}. Once again these SEDs were normalized to an IMF of $1 \msun$ that made it straightforward to scale the flux by the mass of the SPs.

\section{Results}\label{SEC:res}
We analyze our simulations with a focus on Pop III stars in both aggregate and within the context of galaxies, focusing on $7 \le z \le 15$. Figure \ref{fig:sfrd3} depicts the star formation rate density (SFRD) for our simulations, along with observational data compiled by \cite{2014ARA&A..52..415M} and \cite{2016PASA...33...37F}.  Our SFRDs are in reasonable agreement with the LF-based SFRD described by \cite{2016PASA...33...37F} over the range $7 \le z \le 8$ when considering both Poisson error and sample variance \citep{2008ApJ...676..767T}. To minimize parameter differences between runs we use the same star forming efficiency for all runs. We include the 1$\sigma$ estimated errors only for the fid and P3SN runs since they show the greatest variation in the SFRD between simulations. For a detailed discussion on error estimation see \citetalias{2018ApJ...854...75S}.   

Figure \ref{fig:sfrd3} also depicts the Pop III SFRD as well as the `classical' Pop III SRFD that does not include the effects of our subgrid model. The ``Pop III'' rate uses $P_{\star} \times M_{\star}$ to compute Pop III stellar mass. The ``Classical Pop III'' rate only considers whole SPs where $\overline{Z_{\star}} < Z_{\rm crit}$ and hence requires $P_{\star} = 1.0$ for a SP`s mass to be included in the classical SFRD. The three runs with different critical metallicities, $Z_{\rm crit}/Z_{\odot} = \{10^{-4}, 10^{-5}, 10^{-6}$\}, do not show much variation in the Pop III rate. 

\begin{figure}[!t]
\begin{center} 
\includegraphics[width=1\columnwidth]{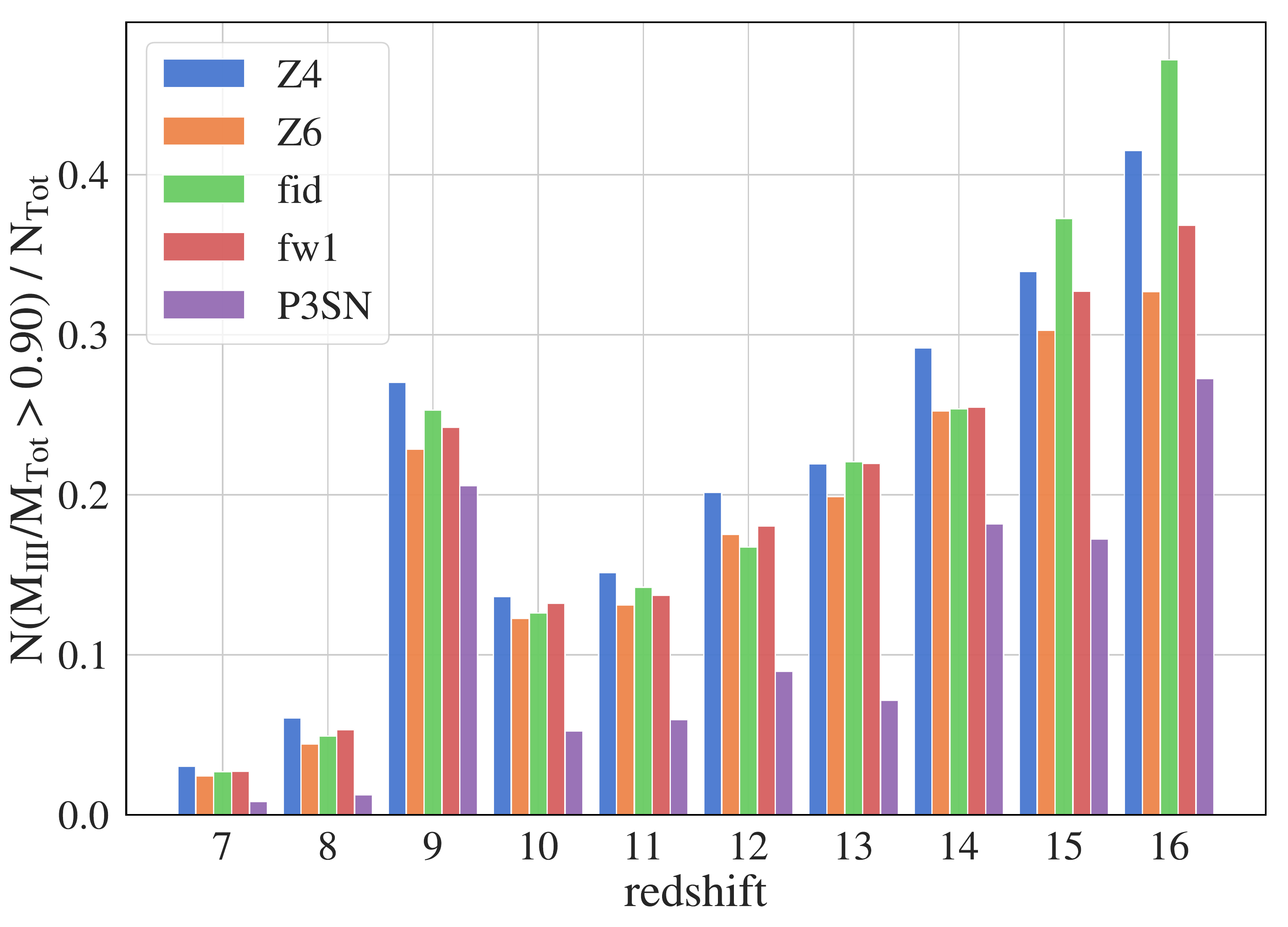} 
\includegraphics[width=1\columnwidth]{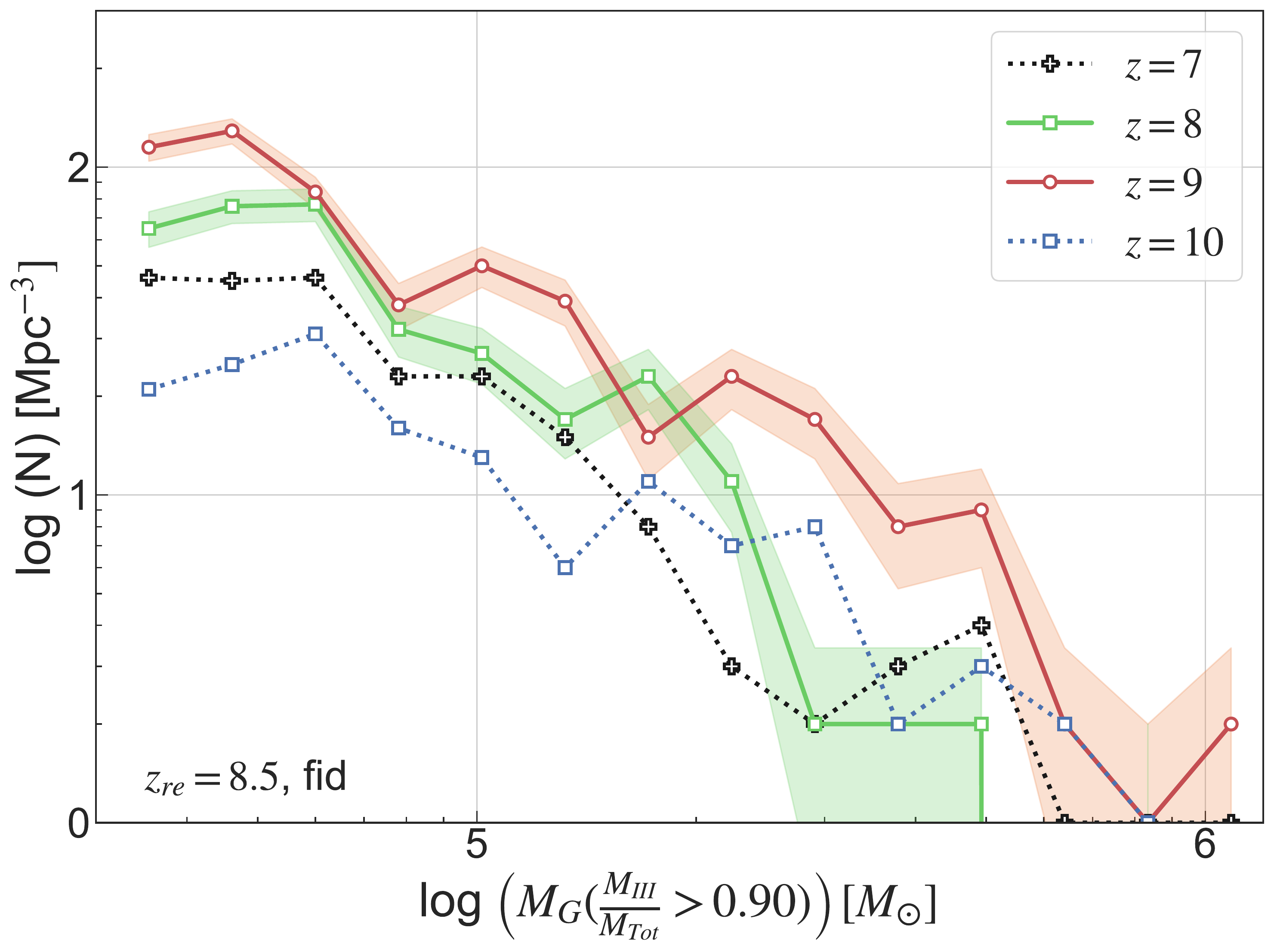}
\caption{\textit{Top:} The fraction of Pop III dominated galaxies (defined as galaxies with at least 90\% of their stellar mass in Pop III stars) across all runs sharply increases immediately before reionization at $z=9$. Considering all of the runs except P3SN, fully 25\% of galaxies, pre-reionization, are Pop III dominated while only 5\% are post-reionization. 
\textit{Bottom:} The histogram depicts counts per comoving Mpc$^3$ for Pop III dominated galaxies, binned by galaxy stellar mass. The plot also indicates that galaxies attain their peak masses, for the fid run, at $z=9$ (red) vs post-reionization (green). The shaded regions indicate 1$\sigma$ Poisson errors. }\label{fig:p3haloStats}
\end{center}
\end{figure}

However, the variation due to modeling the evolution of the pristine gas fraction, $P$, accounts for an average increase of a factor of $\approx$2 (0.3 dex) in the Pop III SFRD, at $z<18$, as compared to the classical rate, when averaged across all five runs. This demonstrates that modeling the subgrid unpolluted fraction of gas is more important than knowing the critical metallicity when attempting to predict the Pop III star formation rate.

Comparing the Pop III SFRDs, the greatest change is caused by the change to the Pop III IMF in run P3SN. This is not surprising since the log normal IMF adopted for Pop III stars results in $\approx$10x more SN energy and pollutants, as well as $\approx$ 90\% less mass in surviving Pop III SPs after 10 Myr, as compared to the other runs. The fid run's Pop III SFRD is approximately 4.3$\times$ that of the P3SN at $z < 18$ although we note that the Pop III SFRD nearly reaches the fiducial level at $z=9$. The P3SN simulation demonstrates how feedback and the IMF are intimately linked, making high-redshift predictions of Pop III stellar populations difficult without further observational data.

\begin{figure*}[!ht]
\begin{center}
\includegraphics[width=.478\textwidth]{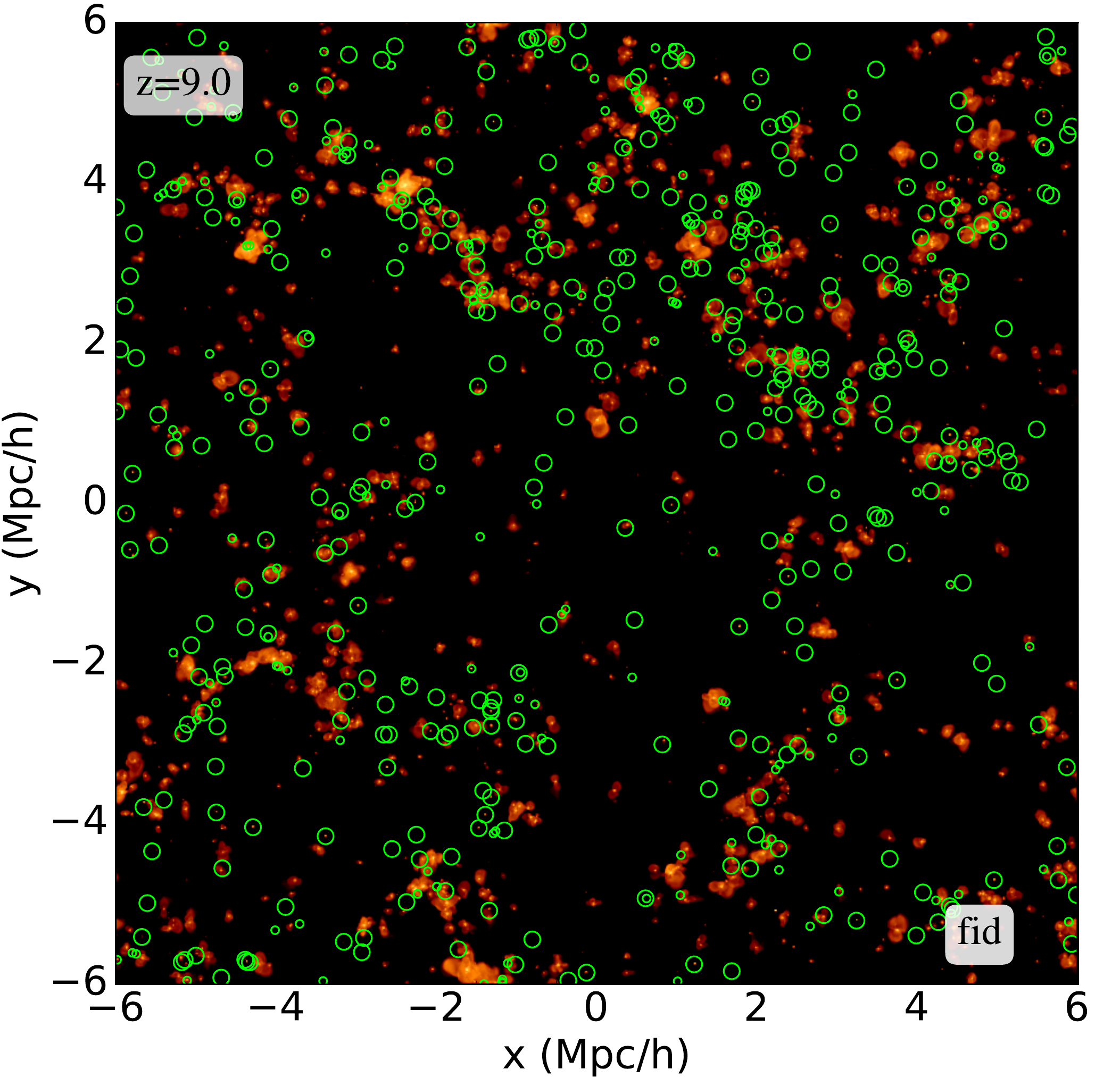}
\includegraphics[width=.513\textwidth]{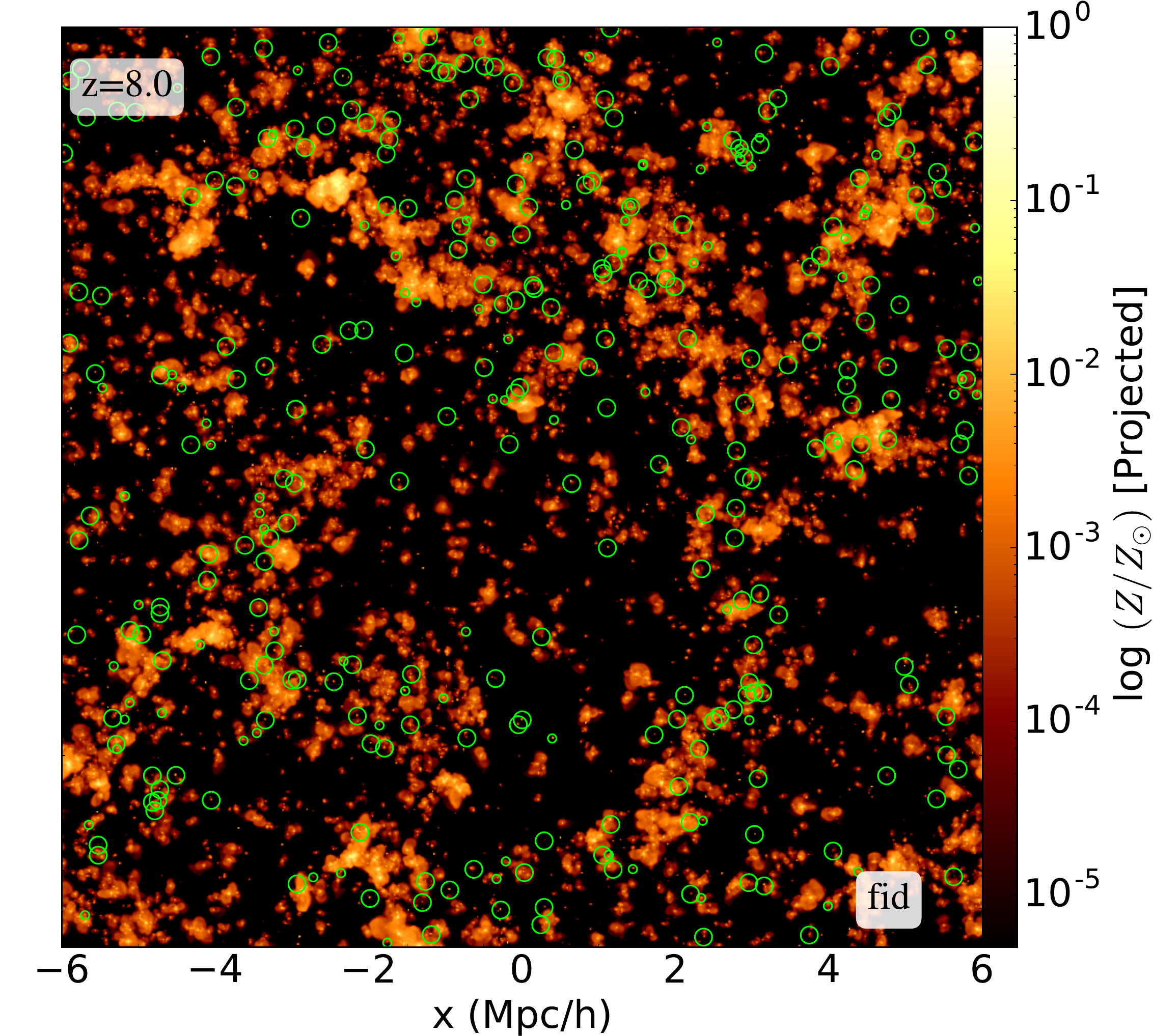} 
\caption{The projected metallicity of the gas for the fid run at $z=9$ (left) and 8 (right).  Circles indicate Pop III dominated galaxies: galaxies with $\ge 90\%$ of their mass in Pop III stars. Circle size indicates the relative mass of each galaxy. At $z=9$, 25\% of galaxies are Pop III dominated and most form in pristine gas away from existing galaxies.  By $z=8$, Pop III star formation in new halos is mostly quenched since reionization raises the gas temperature and therefore the halo mass required to initiate star formation. At this redshift 5\% of galaxies are Pop III dominated. Scale is comoving Mpc h$^{-1}$.\label{fig:p3brightGasMass}}
\end{center}
\end{figure*}

Considering the P3SN run, we also note that the classical Pop III rate falls to zero during several cosmic epochs: $z=18\rightarrow 16,\, 14\rightarrow 12.5,\, 12\rightarrow 11,$ and $10\rightarrow 9$. This is due to the increased metal generation coupled with the instantaneous mixing assumption that is typically used in simulation cells polluted with SN ejecta. Our subgrid model exhibits continued Pop III star formation in regions of unpolluted gas during these intervals again pointing to the improvement in accuracy gained by modeling the mixing time at subgrid scales.

As pointed out in \citetalias{2018ApJ...854...75S}, the increase in star formation immediately before reionization, $z_{\rm re}\approx 8.5$, correlates strongly with an even larger relative increase in the Pop III star formation rate. This is caused by a significant number of new halos crossing the star formation density threshold immediately before reionization increases the temperature of the gas \citep{2008ApJ...682...14F}.  Reionization is essentially an instantaneous event in the simulation. Post-reionization the mass threshold for Pop III star formation in new halos is increased because of the increase in the gas temperature. Since Pop II star formation takes place in more massive galaxies, overall star formation continues at the previous rate, while Pop III star formation drops rapidly. Hence we predict that the largest number of galaxies dominated by Pop III stars -- and Pop III flux -- will be found just prior to reionization.

Figure \ref{fig:p3haloStats}, top, supports this conclusion and depicts the fraction of halos that have at least 90\% of their stellar mass in Pop III stars -- henceforth `Pop III dominated' galaxies -- as a function of redshift. The increase in the number of new Pop III dominated galaxies is clearly visible at $z=9$.  The lower panel depicts a histogram of Pop III dominated galaxies binned in mass, for the fid run, and clearly shows that Pop III dominated halos also attain their highest masses during the pre-reionization epoch.

The top panel in the figure also indicates that $\approx$25\% of galaxies are Pop III dominated, at $z=9$, except in the P3SN case.  However, by $z=8$ the majority of star formation is taking place in pre-existing polluted galaxies and the fraction of Pop III dominated galaxies drops to $\approx$5\%.  While the P3SN run generates the lowest fraction of Pop III stars, the difference between the number of Pop III dominated galaxies pre and post-reionization is just as dramatic. At $z=9$, $\approx21\%$ of galaxies are Pop III dominated. This falls to slightly more than 1\% by $z=8$, again a 20\% drop as exemplified by the other runs.

While the overall fraction of Pop III dominated galaxies is greater at $z > 13$ than at $z = 9$, as we discuss in the next section, the fraction of potentially observable galaxies falls off quickly at $z > 10$ for all of our models.

Figure \ref{fig:p3brightGasMass} depicts the metallicity of the gas for the fiducial run at two redshifts, before and after reionization, along with galaxy locations for Pop III dominated galaxies. At $z=9$, the majority of Pop III dominated galaxies are found in unpolluted regions away from existing galaxies. At $z=8,$ Pop III star formation has dropped by a factor of $\approx$ 5 since reionization has raised the halo mass required to initiate star formation \citep{1986MNRAS.221...53C, 2013MNRAS.432.1989S, 2018arXiv180210096B}. Hence, Pop III star formation occurs mostly in unmixed regions within and on the edges of more massive galaxies.

\begin{figure*}[t]
\begin{center}
\includegraphics[width=.8\textwidth]{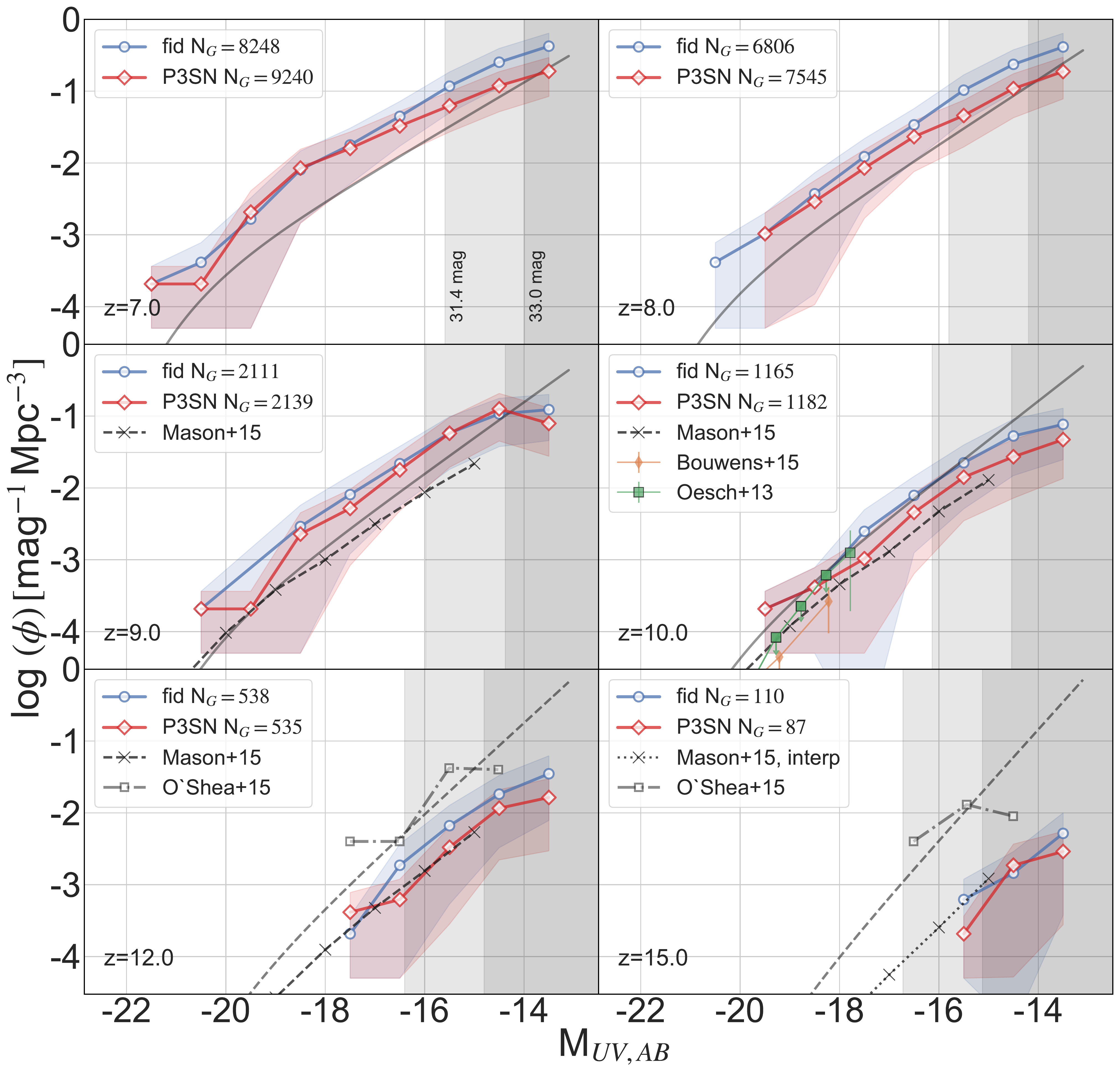}
\caption{UV LFs derived from our P3SN and fid simulations with 1$\sigma$ error bounds including both Poisson noise and sample variance. Magnitude bins are labeled at their right edge. Solid grey lines are \cite{2016PASA...33...37F} Schechter fits. Dashed grey lines are extrapolated. The largest variation between simulations for the parameters studied is between the these runs. P3SN exemplifies a slight drop in the luminosity at the faint end at most redshifts due to the increased SN feedback that helps to quench star formation in less massive halos coupled with an IMF with far fewer long lived, low mass stars. For $z=10$, we have included \cite{2015ApJ...803...34B} and \cite{2013ApJ...773...75O} observational data, with errors. We also include LFs derived from models and simulations by \cite{0004-637X-813-1-21} and \cite{2015ApJ...807L..12O}. We note that for $z=15$ we interpolated the \cite{0004-637X-813-1-21} data that was available for $z=14$ and $z=16$. The shaded areas of the plots indicate likely limiting intrinsic magnitudes for JWST ultra-deep and lensed observations. }
\end{center}
\label{fig:lfFidP3sn}
\end{figure*}

\subsection{Galaxy Luminosity}\label{lc}
Next, we evaluate the effect of our parameters on galaxy flux.  We note that relatively small changes in the fraction of Pop III stars with ages $< 3.5$ Myr can make a significant change to a galaxy's luminosity. 

As noted in Section \ref{SEC:gsm} we used a Salpeter IMF to map SP mass to stars when generating the SED for Pop III, and Pop II, SPs for runs fid, fw1, Z4, and Z6. 
Run P3SN used the log normal IMF SED to model Pop III SP fluxes, with $\eta_{\rm SN, III} = 0.99$, and the Salpeter IMF SED for Pop II SPs, with $\eta_{\rm SN} = 0.10$. The fid\_orig run used a log normal IMF SED to model all Pop III stars even though a Salpeter IMF was assumed in \textsc{ramses} for the SN mass-fraction.

Figure \ref{fig:lfFidP3sn} depicts LFs for two representative simulations across the redshift range $7 \le z \le 15$. Runs Z4, Z6 and fw1 produced LFs nearly identical to fid. The shaded areas indicate the two intrinsic magnitude limits JWST is expected to be able to detect in the deep campaign, $m_{\rm UV} = 31.4$ mag, and via lensing, $m_{\rm UV} = 33$ mag \citep{2006SSRv..123..485G}.

The two runs produce essentially indistinguishable LFs when considering 1$\sigma$ estimated errors based on Poisson noise and sample variance. Even the run P3SN, that generates $\approx10\times$ the SN feedback as compared to the other runs, displays only a small change in galaxy luminosities across the redshift and magnitude range depicted. However, the combination of feedback and the reduced number of surviving Pop III stars for run P3SN does slightly reduce the brightness of galaxies. 

For comparison, we include LFs by \cite{0004-637X-813-1-21} and \cite{2015ApJ...807L..12O}. The former work uses a semi-analytic framework to model the UV LF based on the star formation history in DM halos. While our LF is slightly higher at $z=9$, we are in good agreement with their model at $z\geq 10$.

The work by \cite{2015ApJ...807L..12O} is based on the Renaissance simulations. At $z\geq 12$, where comparisons are available, we underperform their simulation. However, we note that the Renaissance simulations best resolution is 19 pc as compared to our 64.5 pc, which allows them to capture star formation earlier, in smaller halos.

The consistency of these result across simulations indicates that our predictions for the luminosity of high redshift galaxies in \citetalias{2018ApJ...854...75S} are robust across this parameter space. While we have yet to look at the effects of self-consistently treating reionization, it is promising to note that the luminosity of these galaxies does not seem to be highly correlated with the critical metallicity, SN loading or the additional energy generated by a top-heavy Pop III IMF.

\subsection{Pop III Flux}\label{p3f}
While our parameter variations do not result in striking changes to the LFs of our high redshift galaxies, they do produce changes in the fraction of Pop III flux coming from them.  
Here we briefly discuss galaxies with a Pop III flux fraction down to $10^{-3}$ but focus on `observable Pop III-bright galaxies'  that have $m_{\rm UV} \le 31.4$ mag  and  $\nicefrac{f_{\rm III}}{f_{\rm Tot}} \ge 0.75$. These  galaxies are important to future observational searches for Pop III stars. 

\begin{figure*}[t]
\begin{center}
\includegraphics[width=0.95\columnwidth]{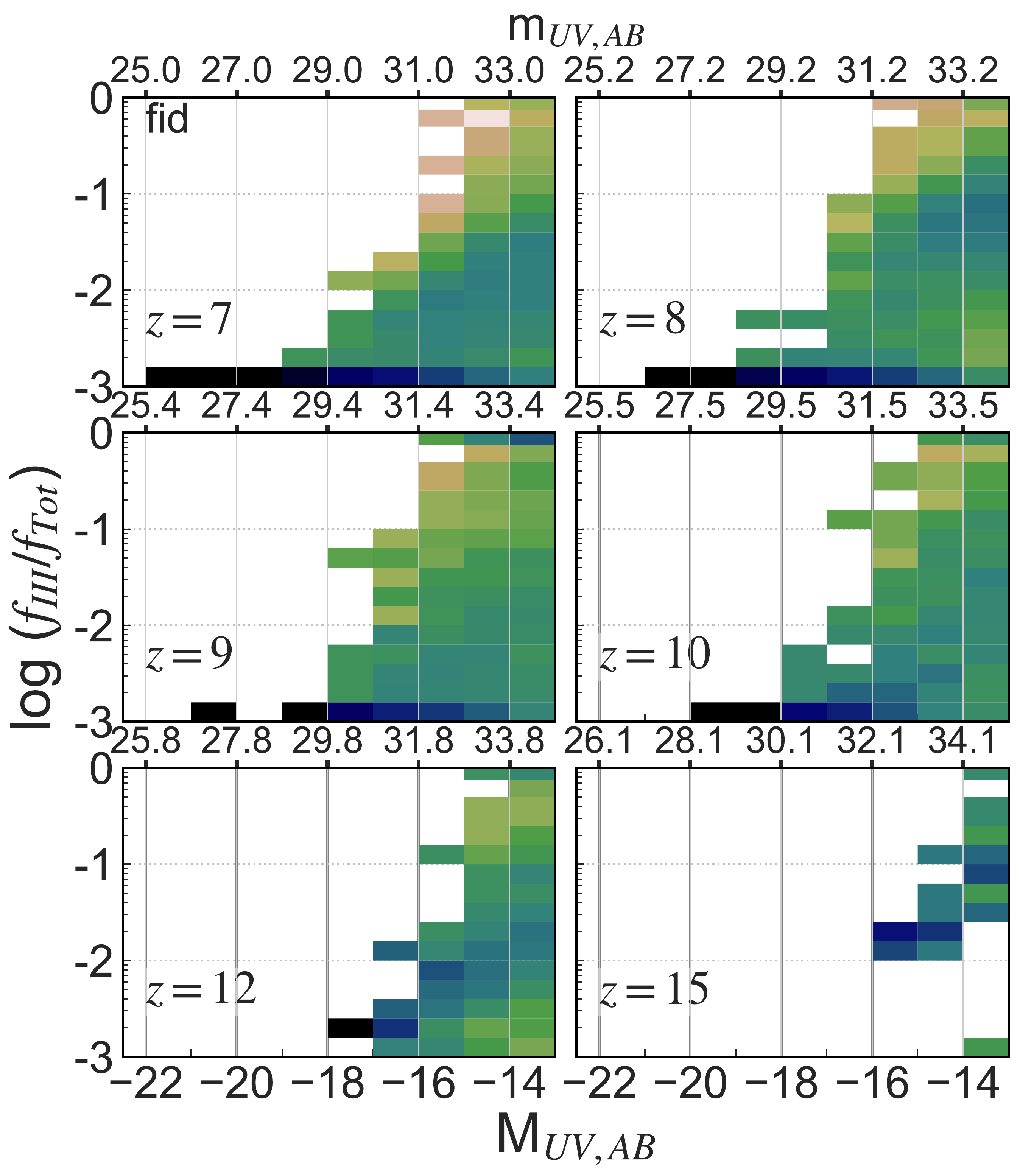} 
\includegraphics[width=1.055\columnwidth]{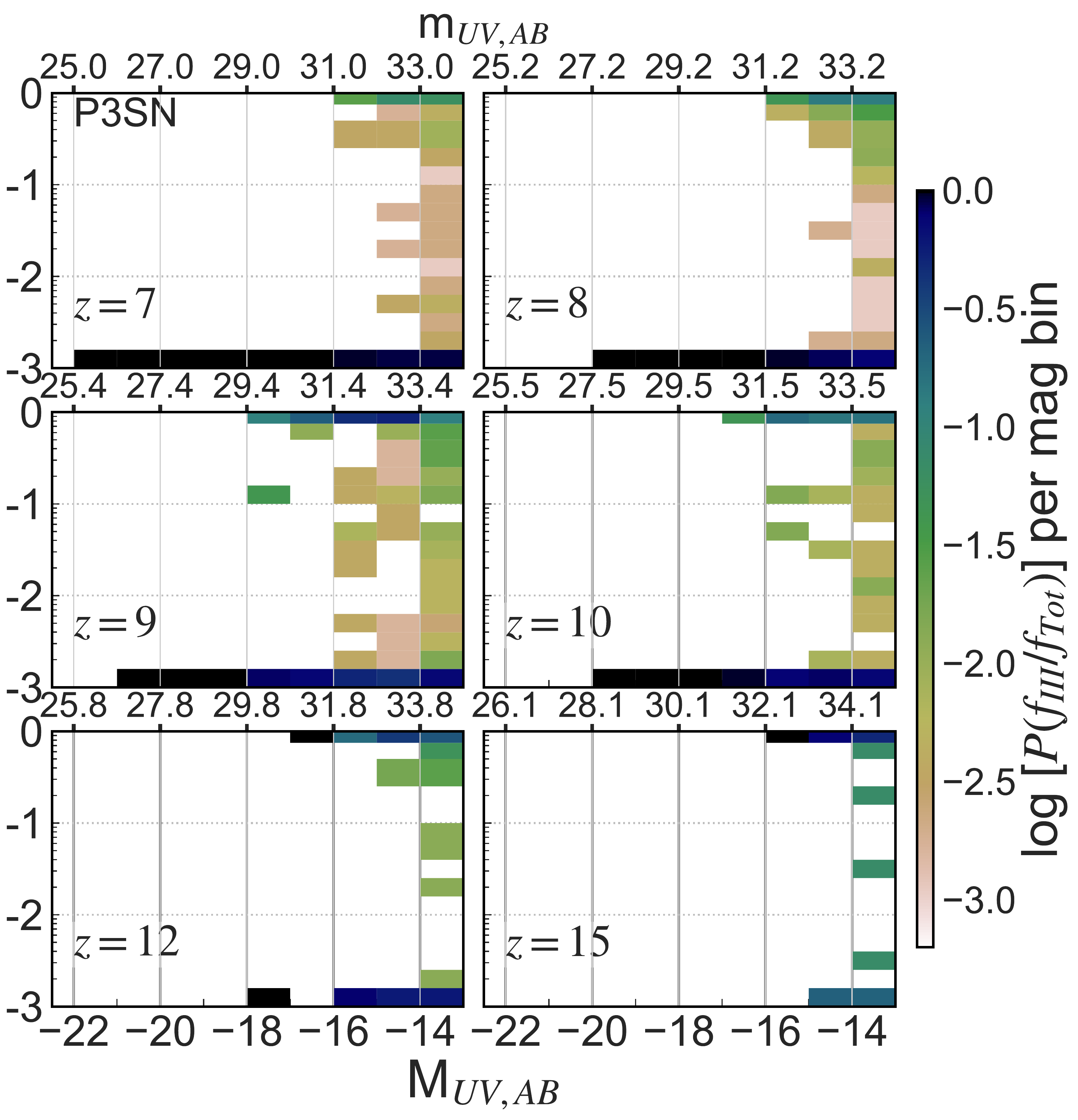} 
\caption{The normalized probability, per magnitude bin, for the rest-frame UV Pop III flux fraction, $\nicefrac{f_{\rm III}}{f_{\rm Tot}}$, for our galaxies. Magnitude bins are labeled at their right edge. The top row of bins represent $\nicefrac{f_{\rm III}}{f_{\rm Tot}} \ge 75\%$, the next row, 50\%.  \textit{Left:} The fid run that used a Salpeter IMF for all SPs. Here, Pop III stars are low mass and longer lived, giving rise to hybrid galaxies with small fractions of Pop III flux. \textit{Right:} Run P3SN. Here the probabilities of finding hybrid galaxies with $10^{-3} < \nicefrac{f_{\rm III}}{f_{\rm Tot}} < 0.5$ decreases drastically. This is due to the increased feedback in the P3SN simulation combined with the lack of older Pop III stars. However, the fraction of Pop III-bright galaxies (top row of bins) is greater than fid at all redshifts since Pop III star formation in new halos that follows the log normal IMF results in very luminous high-mass stars. While the fraction of galaxies with $\nicefrac{f_{\rm III}}{f_{\rm Tot}} \ge 0.75$ is significant at $z\ge9$, the vast majority of galaxies are beyond the magnitude limits for unlensed JWST observations. See Figure \ref{fig:p3brightFrac} which summarizes the data for Pop III-bright galaxies. \label{fig:p3Prob2}}
\end{center}
\end{figure*}

Figure \ref{fig:p3Prob2} depicts the Pop III flux fraction for runs fid and P3SN that display the greatest difference across redshifts.  We have normalized the probabilities in each magnitude bin such that they sum to one making it easy to see the distribution of Pop III flux per magnitude bin. For the fid run, Pop III flux from hybrid galaxies, $10^{-3} < \nicefrac{f_{\rm III}}{f_{\rm Tot}} < 0.5$, is dominant since Pop III stars are assumed to be lower mass and hence long-lived. For P3SN this pattern is inverted: we note a dearth of hybrid galaxies since massive Pop III stars from the log normal IMF live short, but radiatively intense lives. This results in relatively little Pop III flux except for the small fraction of galaxies in which these massive stars are shining. Here, they dominate the flux of the galaxy (top row of bins). However, once they are gone the Pop III flux quickly falls below $\nicefrac{f_{\rm III}}{f_{\rm Tot}} \le 10^{-3}$. This bimodality is most evident before reionization, when new halos form Pop III stars from the pristine gas. Once they die the galaxy is dominated by Pop II flux. However, most of the Pop III-bright galaxies at $z\ge9$ are on the edge of visibility for giant telescopes like JWST. At lower redshift, $z < 9$, the vast majority of galaxies have only a tiny fraction of Pop III flux: $\nicefrac{f_{\rm III}}{f_{\rm Tot}} \approx 10^{-3}$, agreeing with the lack of current observations of Pop III-bright galaxies.

\begin{figure}[!ht]
\begin{center}
\includegraphics[width=1\columnwidth]{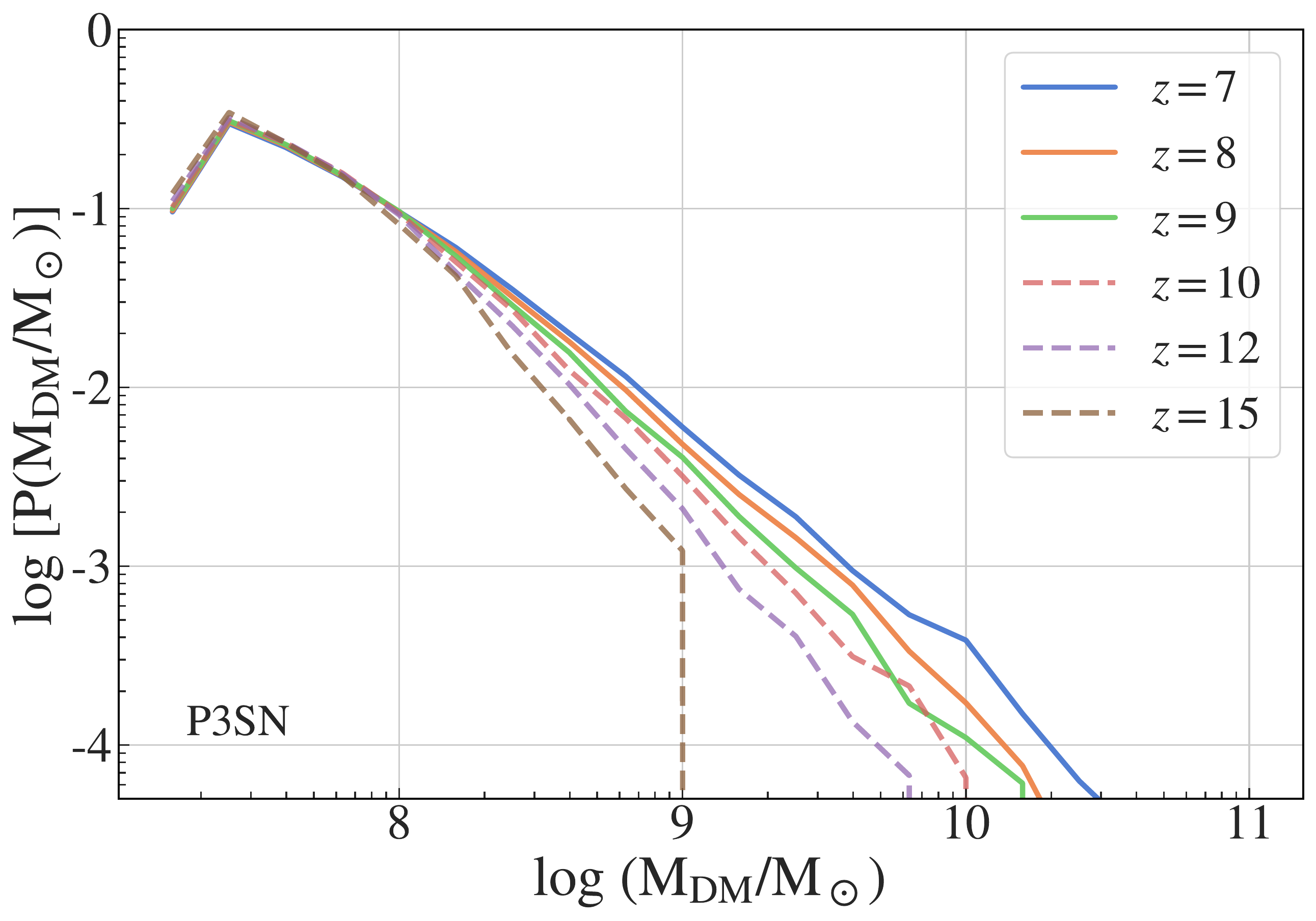} 
\includegraphics[width=1\columnwidth]{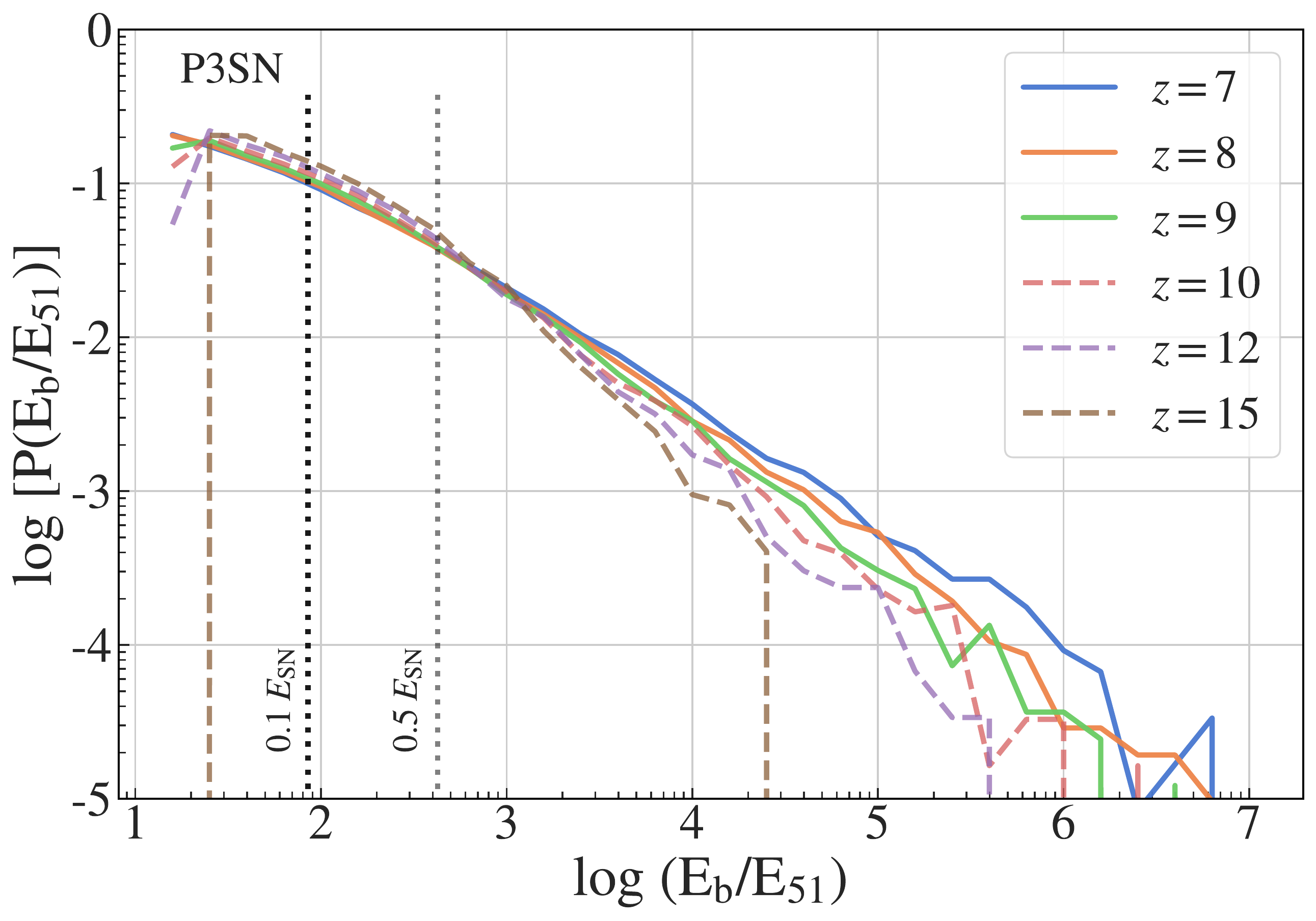}
\caption{\textit{Top}: The halo DM mass function (HMF) for P3SN expressed as a normalized PDF. As expected, the HMF indicates that the majority of galaxies have $M_{\rm DM} < 10^{8} \msun$ at all redshifts. \textit{Bottom}: The gravitational binding energy (in units of $10^{51}$ erg) for P3SN galaxies expressed as a normalized PDF. The dotted line labeled `0.1 $E_{\rm SN}$' indicates a conservative fraction of SN energy (10\%) that couples to the gas when considering the minimum SP mass. Galaxies with binding energies below this threshold should lose their baryons as a result of a single P3SN Pop III SP SN. This includes galaxies up to the median binding energy. An average of 67\%, by number, of halos become unbound by P3SN SN, across redshifts. We include a higher coupling efficiency factor of 0.5 $E_{\rm SN}$, for reference.}
\end{center}
\label{fig:hmrbe}
\end{figure}

The second factor contributing to the bimodal flux distribution in run P3SN is the 10 fold increase in SN energy. SNe in P3SN efficiently evacuate the gas from the host halo contributing to the reduction in the number density of hybrid galaxies.  Considering the typical (also the minimum) SP mass, $M_{\star} \ge 8.3 \times 10^{3}\, \msun$, we find the SN energy generated by Pop III SNe is at least $8.2\times 10^{53}$ erg. 
Figure \ref{fig:hmrbe}, top, depicts the range of halo DM masses in this simulation, as a PDF, along with a plot of the estimated gravitational binding energy \citep{2010hdfs.book.....L}, 
\begin{equation}\label{eq:haloBind}
\begin{aligned}
E_{\rm b}=2.9\times 10^{53} \left(\frac{M_{\rm h}}{10^{8}\msun}\right)^{5/3}\,\left(\frac{1+z}{10} \right)\, \rm{erg},
\end{aligned}
\end{equation}
bottom, for these halos. When considering a SN to gas coupling efficiency of 10\% \citep{2018MNRAS.478.1795H, 2005ApJ...630..675K, 2008ApJ...682...49W}, we note that Pop III SNe efficiently evacuate the gas from median mass halos and below, which have binding energies $\lesssim 4.3\times 10^{52}$ erg.

Our stellar and SN feedback models are not directly comparable to those described in the above references\footnote{\cite{2018MNRAS.478.1795H} modeled a coupling efficiency of 10\% between the SN energy and the surrounding gas. Our simulations assumed that 10$\times$ the mass ejected by SN is swept up and carried out of the cell.} However, our purely mechanical SN feedback acts on SN ejecta mass on the order of the SP mass for our P3SN simulation. This level of parameterized coupling effectively helps to evacuate gas and quench star formation in SN-hosting cells.

\begin{figure}[!ht]
\begin{centering}
\includegraphics[width=.995\columnwidth]{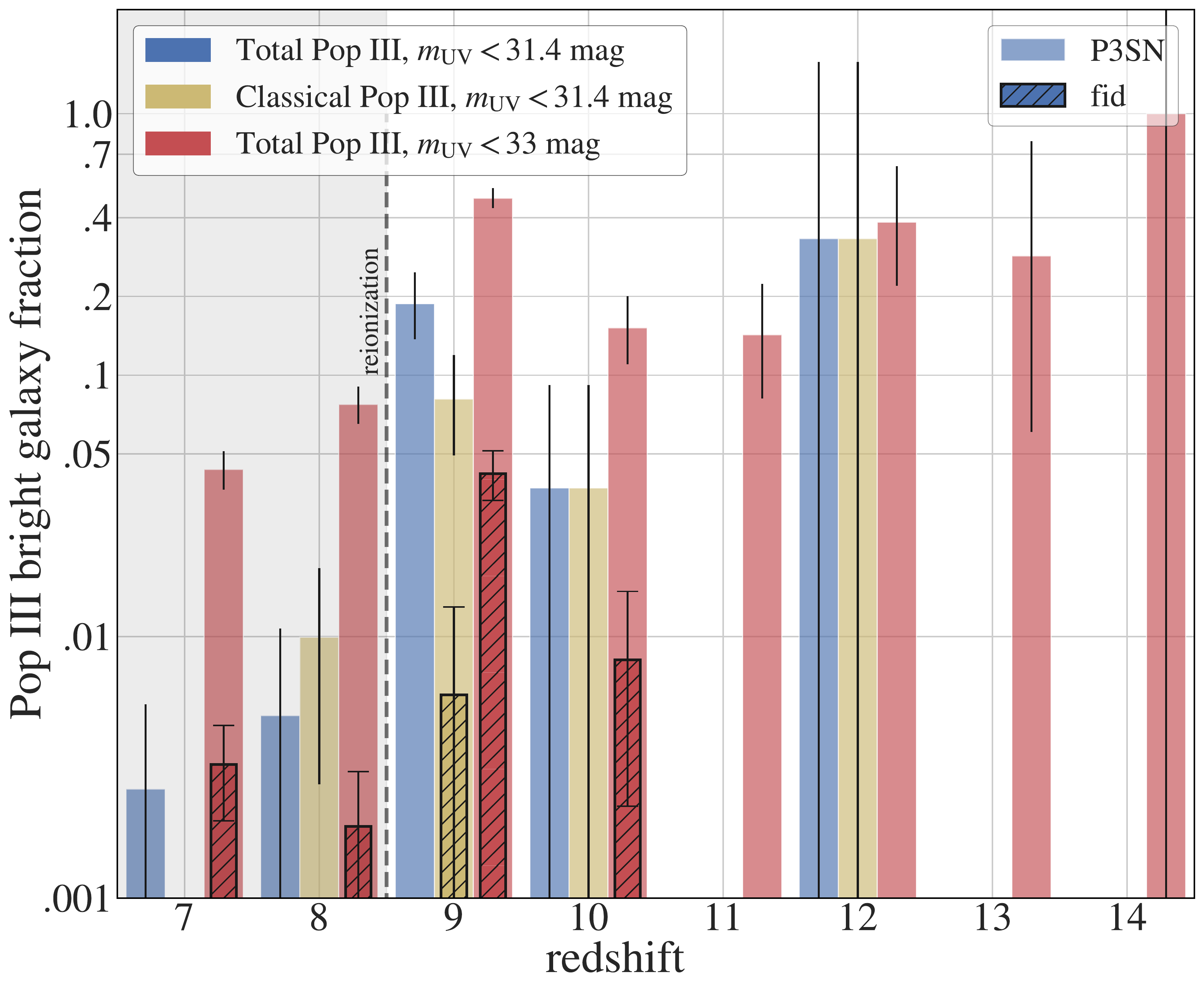} 
\caption{A comparison of the fraction of Pop III-bright galaxies for the P3SN and fid runs, with Poisson error bars. The plot depicts the probability, per redshift, of finding an observable, $m_{\rm UV} \le 31.4$ mag (blue) or $m_{\rm UV} \le 33$ mag (red), Pop III-bright, $\nicefrac{f_{\rm III}}{f_{\rm Tot}} \ge 0.75$, galaxy as a fraction of all galaxies meeting the magnitude cutoff. We include yellow bars that consider only classical Pop III star formation for comparison. For P3SN, 19\% of observable galaxies at $z=9$ are Pop III-bright. At $z=12$, 33\% are Pop III-bright galaxies but there are only 3 galaxies with $m_{\rm UV} \le 31.4$ mag resulting in a large uncertainty. We conclude the best epoch to look for observable Pop III-bright galaxies is immediately before reionization. \label{fig:p3brightFrac}}
\end{centering}
\end{figure}

Pop III-bright galaxies, as a fraction of all observable galaxies, are summarized in Figure \ref{fig:p3brightFrac}. Since the fractions of Pop III-bright galaxies for runs fw1, Z4 and Z6 are very similar to the fid run, here we only include a comparison between fid and P3SN. For a comparison between fid and fid\_orig see the \appendixname{}. We use the JWST limiting magnitude of  $m_{\rm UV} = 31.4$ mag, along with a lensing magnitude limit of $m_{\rm UV} = 33$ mag to identify observable galaxies with at least 75\% of their flux coming from Pop III stars. 

For run P3SN the fraction of observable Pop III-bright galaxies is 33\% at $z=12$. However, there are only 3 galaxies with $m_{\rm UV} \le 31.4$ mag at this redshift and hence the uncertainty is very large. The statistics are similarly weak at $z=10$ where we find a single Pop III-bright galaxy out of 27 observable galaxies: 4\%. Hence, we note that, as in our previous work, the largest statistically significant fraction, $\approx19\%$, of observable Pop III-bright galaxies occurs immediately before reionization. Again, this is due to new halos crossing the density threshold for star formation before the UV background raises the temperature of the gas. 

As shown when discussing Figure \ref{fig:p3Prob2}, none of our galaxies were Pop III-bright at the $m_{\rm UV} = 31.4$ mag limit for the fid run. To detect galaxies with a significant fraction of Pop III flux we have to go to $m_{\rm UV} = 33$ mag where, once again, the largest statistically significant fraction of Pop III-bright galaxies occurs at $z=9$. The flux from the fid run's Salpeter IMF of longer-lived, low mass stars simply does not out-shine the few, short-lived massive stars modeled by the log normal IMF in run P3SN.

Once again, reionization in our simulation is essentially an instantaneous event. In reality it was a patchy, extended process and hence our result does not map universally onto any observed field where sample variance is a factor.

The greatest effect on predicted counts of Pop III-bright galaxies occurs when we consider a JWST lensing limit of $m_{\rm UV} \le 33$ mag and run P3SN.  At $z=10$ and 11 we predict $\approx$15\% of Pop III-bright galaxies will have an intrinsic magnitude such that 10$\times$ lensing magnification will make them observable by the JWST. At $z=12$, run P3SN has approximately the same fraction of $m_{\rm UV} \le 33$ mag Pop III-bright galaxies as at $m_{\rm UV} \le 31.4$ mag, but the Poisson uncertainty for the former is far lower. While the $z=13$ and 14 statistics also indicate significant fractions of Pop III-bright galaxies at $m_{\rm UV} \le 33$ mag, the 1$\sigma$ uncertainties are large. However, even considering the worst case, more than 5\% of Pop III-bright galaxies at $z=13$ have $m_{\rm UV} \le 33$ mag. Hence, lensing campaigns should provide JWST with opportunities to observe Pop III galaxies beyond $z=10$, but once again the redshift immediately before reionization contains the largest fraction of Pop III-bright galaxies when considering Poisson statistics.

\subsection{Chemical Composition}\label{chem}
Using $\overline{Z_{\rm P, \star}}$, $\overline{Z_{\star}}$, and $P_{\star}$ we model the fraction of stellar mass in each SP that represents Pop III stars as well as the enhanced metallicity of the polluted fraction of Pop II stars, $Z_{\star} = \overline{Z_{\star}}/(1-P_{\star})$ as described in Section \ref{sec:cm}. To simplify direct comparisons with our earlier work, we once again, for our initial analysis, adopt the elemental abundance pattern generated by a $60\, \msun$ Pop III SN \citep{hegerwebsite} as representative of metal yields in the $20 \msun \leq M_{\star} < 120 \msun$ range, the dominant region of our Pop III IMF that produces carbon enhanced material. Specifically, the abundances of each element in each SP are computed by mapping the mass fraction of primordial metals, $Z_{\rm P, \star}$, to the abundances found in this ejecta. `Regular metals', that have a mass fraction $Z_{\star} - Z_{\rm P, \star}$, are modeled using abundances provided by \cite{Timmes} and are representative of typical Type II SN abundances. Details are discussed in \citetalias{2017ApJ...834...23S}.

Note that the primordial metals, $Z_{\rm P, \star}$,  were generated by the mass fraction of each Pop III SP that underwent a SN after 10 Myr. These are stars $\gtrsim 16 \msun$ in each IMF. This corresponds to 99\% of the mass of each SP for our log normal IMF but only 10\% of the SP mass for the Salpeter IMF.

Using the criteria from \cite{2005ARA&A..43..531B}, we identified the mass fractions of our Pop II SPs that are ultra metal poor (UMP, [Fe/H] $< -4.0$) and carbon enhanced ([C/Fe] $> +0.7$). Our fraction of such stars for runs fid, Z4, Z6, and P3SN are in very good agreement with observations for this class of star in the MW halo \citep[see][Figure 6]{2018arXiv180604738Y}. Comparisons to the fraction of observed CEMP-no stars at metallicities greater than UMP diverge with the simulations underperforming observations since stars with [Fe/H] $> -4.0$ are still being formed in the simulation at $z\ge7$.

\begin{figure*}[t]
\begin{centering}
\includegraphics[width=.465\textwidth]{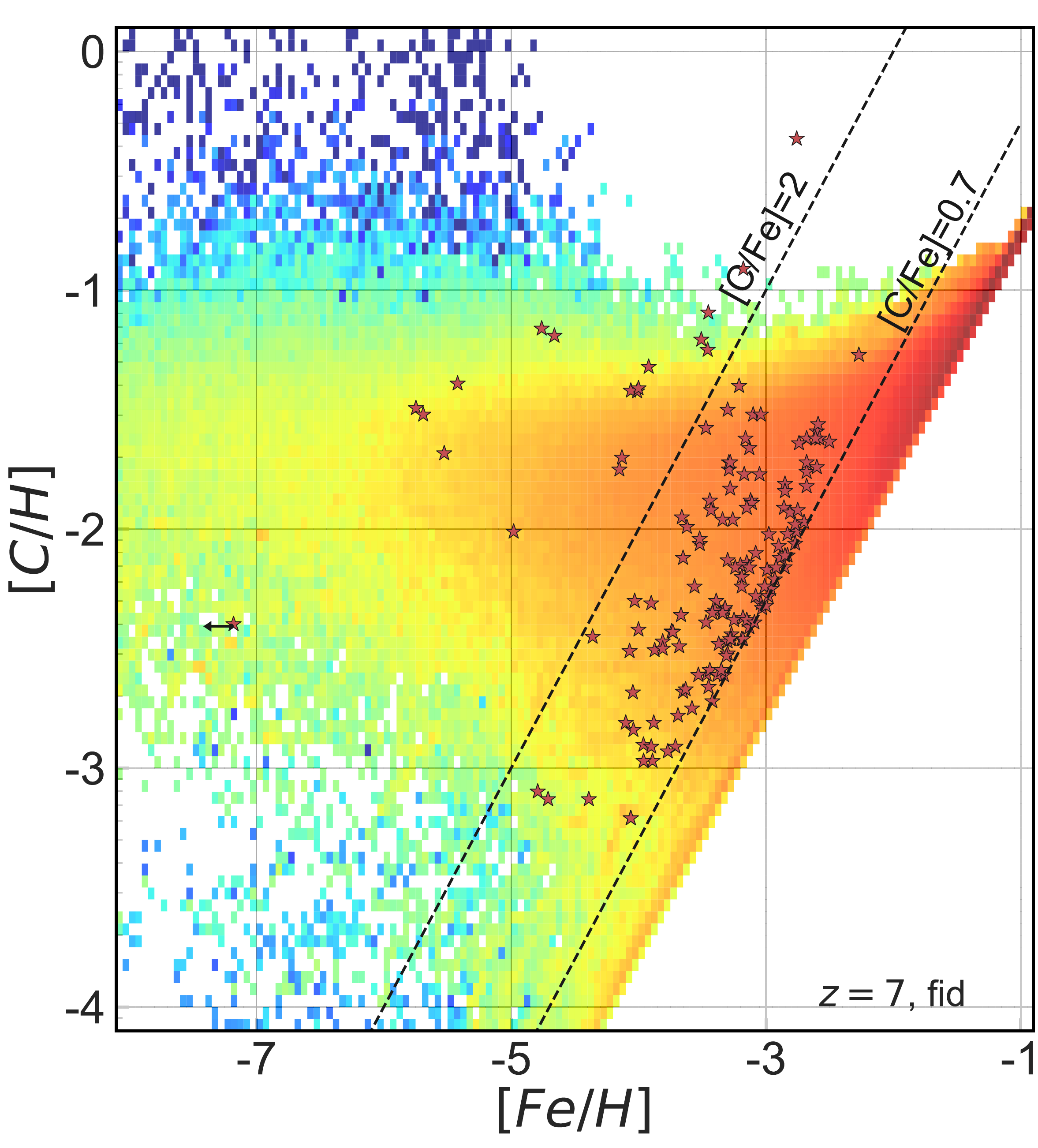} \hspace{-.15cm}
\includegraphics[width=.524\textwidth]{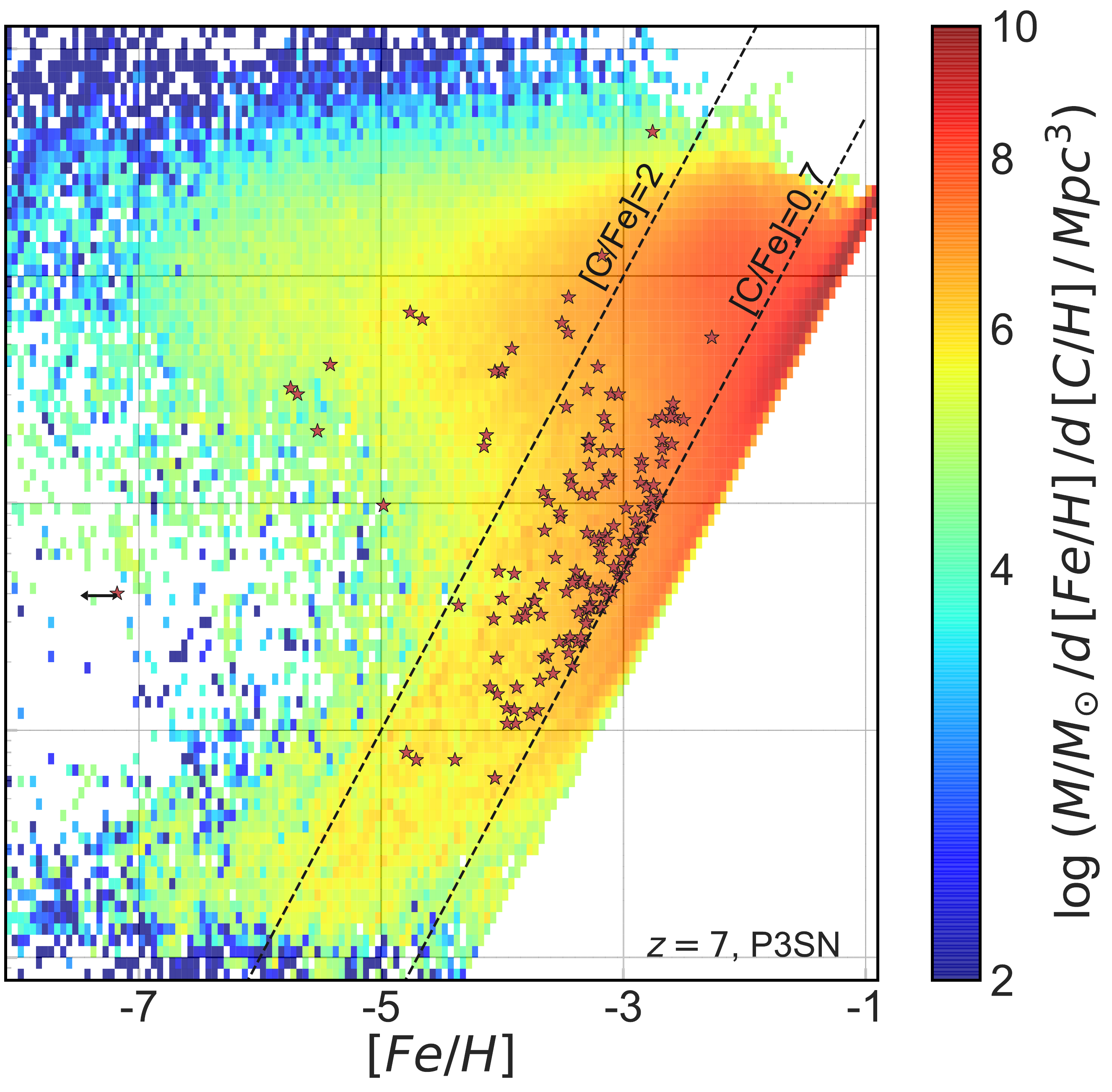} 
\caption{Joint PDFs depicting the mass-weighted probabilities for the chemical abundance of [C/H] as a function of [Fe/H] for all SPs in our fiducial, \textit{left}, and P3SN, \textit{right}, simulations at $z=7$. Carbon levels are higher in the P3SN simulation owing to the larger fraction of carbon-rich Pop III SN ejecta generated by the top heavy IMF. It is the best match to observations of CEMP-no stars \citep[red stars,][]{2016ApJ...833...20Y}. The P3SN plot also depicts an enhanced probability of finding stars with $0 < $~[C/Fe]$\; \lesssim 1$ as well as a population of more MP stars with [Fe/H]$\; < -2$ and [C/H]$\; \approx -1$.  The dashed lines indicates  [C/Fe]$\; = 0.7$ and  [C/Fe]$\; = 2.0$. \label{fig:2dchempdf}}
\end{centering}
\end{figure*}

Moving on to a more detailed chemical analysis, Figure \ref{fig:2dchempdf} depicts the mass-weighted joint PDFs for [C/H] as a function of metallicity for the fid and P3SN simulations' SPs. Results for runs Z4, Z6 and fw1 are very similar to the fid run. We focus on carbon since it characterizes CEMP-no stars \citep{2005ARA&A..43..531B}.  The observed chemical composition of CEMP-no stars in the MW halo requires SN progenitors from a top-heavy IMF with stars that end their lives as Type II SN leaving a BH or neutron star that traps the heavier elements. Several stars in the mass range of our log normal IMF result in SN with such carbon-enhanced yields \citep{2002ApJ...567..532H}.

Each plot is overlaid with the set of CEMP-no stars from \cite{2016ApJ...833...20Y} and includes a dashed line depicting a reference [C/Fe] ratio of 0.7. While both plots depict an enhanced probability of finding stars with [C/Fe] $\;\gtrsim 0.5$, the P3SN data more clearly displays the bimodal relationship between carbon and iron discussed by \cite{2016ApJ...833...20Y}. This plot depicts the enhanced probability of finding stars with $0 < [{\rm C/Fe}] \lesssim 1$, likely indicative of a population of CEMP-no stars polluted by both Pop III and Pop II SN, as well as a population of very metal poor stars with [Fe/H]$\; < -2$ and [C/H]$\; \approx -1$, a population likely polluted solely by Pop III SN.  

\begin{figure*}[t]
\begin{centering}
\includegraphics[width=1.0\textwidth]{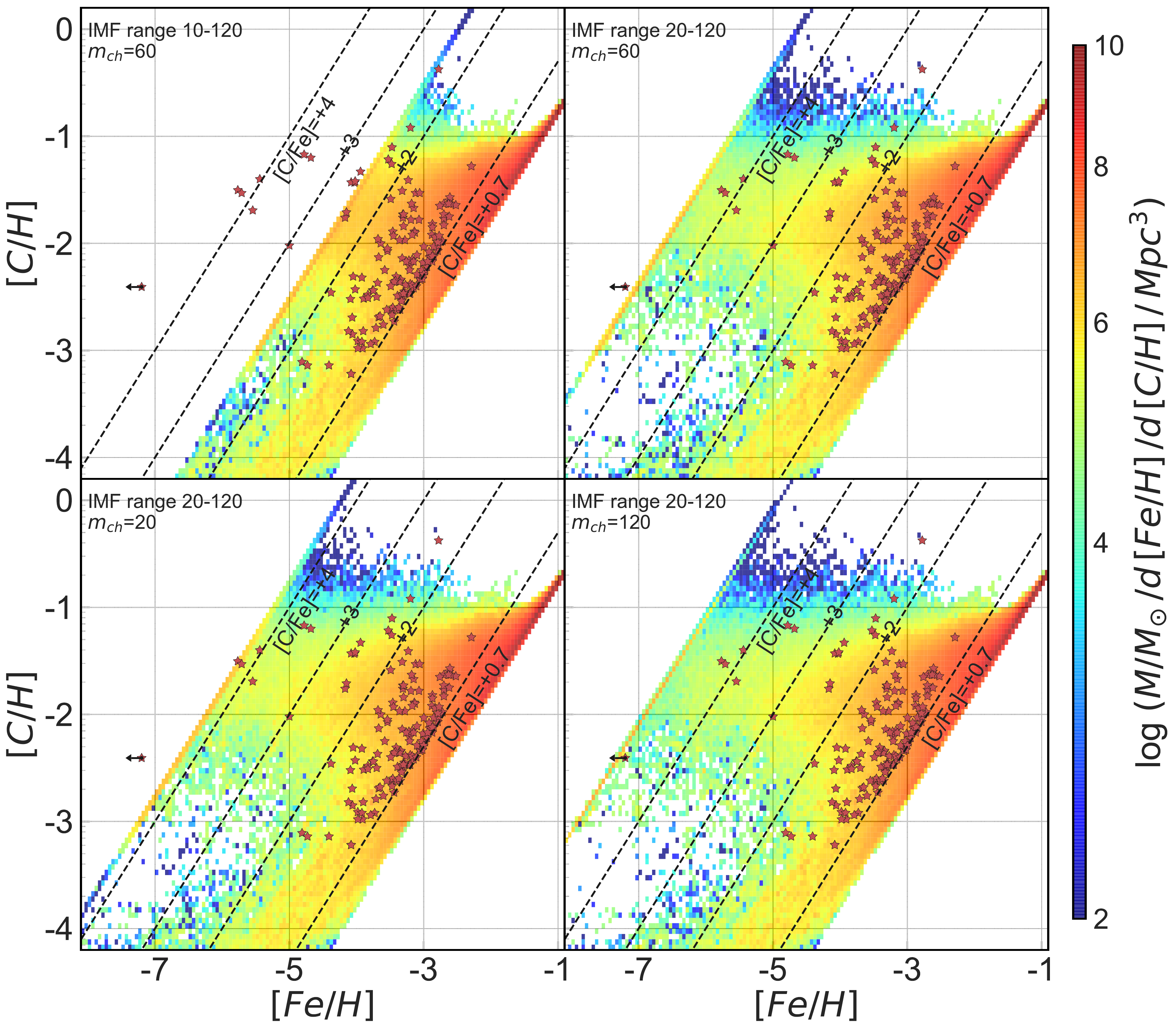} 
\caption{Same as Figure \ref{fig:2dchempdf} but for an integrated IMF of yields over the range specified in each subplot. All data based on run P3SN at $z=7$ and yields from \cite{2010ApJ...724..341H}. The characteristic mass was also varied as indicated. Comparing the panels, we note that adjusting the floor of the mass range has a larger effect on the final ratio of [C/Fe] than the characteristic mass. Once again observations are from \cite{2016ApJ...833...20Y}. \label{fig:cempCompare}}
\end{centering}
\end{figure*}

We also model Pop III SN ejecta integrated over a variety of sub-ranges of our P3SN IMF using the yields from \cite{2010ApJ...724..341H} and \cite{2013ARA&A..51..457N}
making use of the SYGMA \citep{2018ascl.soft06019R, 2018ApJS..237...42R} software library.  

Figure \ref{fig:cempCompare} depicts the results of this analysis. We only include plots for a representative subset of the runs using Pop III SN yields from \cite{2010ApJ...724..341H}. The yields from \cite{2013ARA&A..51..457N} include the iron rich ejecta from PISN (progenitors from $140 \msun$ to 300 $\msun$) and result in subsequent generations of stars with [C/Fe]$\; \approx 0$. Even restricting the \cite{2013ARA&A..51..457N} yields to a maximum progenitor of 120 $\msun$ we find the resulting range of [C/Fe] in our Pop II stars do not match the range of CEMP-no observations as well as the yields generated by \cite{2010ApJ...724..341H}.

The plot in the upper left of Fig. \ref{fig:cempCompare} covers the mass range from 10 to 120 $\msun$, near the top of the range in which heavier elements are trapped in a compact remnant. While Pop III stars with $M \lesssim 16 \msun$ likely live longer than the 10 Myr assumed for Pop III SN, we include their yields in the subplot to demonstrate the effect these lower mass stars have on the [C/Fe] distribution.

Adjusting the low-end mass of the IMF to 20$\msun$, the upper right subplot, results in a better match to the observed chemical abundances of CEMP-no stars since the amount of iron in SN ejecta from $10 \msun$ progenitors lowers the maximum [C/Fe] ratio substantially. Assuming a higher characteristic mass of 120 $\msun$, the bottom right subplot, results in the best match when integrating SN yields across an IMF. However, as can be seen in the lower left panel, the value of the characteristic mass does not have a large effect on the probability distribution. Changing the characteristic mass from 20 $\msun$ to 120 $\msun$ changes the maximum [C/Fe] ratio by 0.67 dex. However, changing the bottom of the IMF range from 10 to 20 $\msun$ results in an increase in the [C/Fe] ratio of $\approx$ 2.1 dex or a factor of $\approx$ 138. Lastly, we note that the overall peak carbon abundances, relative to solar, are lower when integrating an IMF of SN yields as compared to the single progenitor model depicted in Figure \ref{fig:2dchempdf}.

This analysis argues for an IMF biased toward stars with a large characteristic mass and a steep drop in the fraction of stars below 20 $\msun$ -- one that is greater than that modeled by a log normal IMF. Devising and modeling such IMFs could be tackled in a follow-on project. Of course, these results are very sensitive to the adopted set of Pop III yields, and these are still highly uncertain.  We also note that not all halos would generate stars across the entire mass range. If the first minihalos generated small clusters of stars a stochastically sampled IMF would better model the resulting chemical enrichment.

Even so, this agreement between the P3SN simulation and observations further supports the idea of a top-heavy IMF. It ties the nucleosynthetic products of Pop III SN to the chemical composition of ancient MW CEMP-no halo stars.

\section{Conclusions}\label{SEC:con}
We have conducted a parameter study analyzing the results of 5 large-scale cosmological simulations to assess the observational consequences of some of the unknown properties related to Pop III star formation. In particular, our simulations have explored the effects of
\begin{enumerate}
\item lowering the SN mass loading factor, that describes the amount of circumstellar gas carried along with SN ejecta, from 10 times the ejecta mass to 1,
\item varying the value of the critical metallicity that marks the boundary between Pop III and Pop II star formation between $10^{-6}$ and $10^{-4} Z_{\odot}$,
\item implementing a log normal Pop III IMF in place of the Salpeter IMF, resulting in an $\approx$10-fold increase in both the amount of SN energy and metals injected into the ISM.
\end{enumerate}
We have compared these simulations to each other and to the run from \citetalias{2018ApJ...854...75S}, fid\_orig, quantifying the differences and analyzing the implications for predictions made in our previous work.

We find that the parameter range explored does not produce a large spread in the overall SFRD. The largest change, produced by P3SN, results in approximately a 50\% reduction in the total SFRD at $z\le12$ and less at $z \le 9$. The differences between the fid and P3SN run are within sample variance at $z\le12$.  Looking at the Pop III SFRD, run P3SN again produces the largest change from the fiducial case, this time significant, resulting in a decrease of an average of approximately 0.64 dex (a factor of $\approx$1/4) averaged over the redshift range $7 \le z < 18$. This points to the importance of understanding the Pop III IMF since it has a direct impact on the fraction of surviving Pop III stars at every epoch. Additionally, understanding the IMF relates directly to the amount of SN feedback. The amount of SN feedback is important since it is relatively easy to evacuate the gas from low mass halos in the early universe.

However, once again we see that modeling the pristine fraction of gas has a significant effect on the Pop III SFRD. The subgrid model produces a Pop III SFRD a factor of 2 above the classical model for all simulations except P3SN. For P3SN the difference between the classical Pop III SFRD and the subgrid model was even more pronounced with the classical rate falling to zero during several epochs.  We did not see an appreciable difference when varying the critical metallicity for Pop III star formation over 2 orders of magnitude. This reinforces our conclusion that even at a physical resolution of 64.5 pc, modeling the subgrid evolution of the pristine fraction of gas is more important than knowing the value of the critical metallicity.

We find that Pop III star formation peaks immediately before reionization, in all of our simulations. This is true in terms of both the mass of Pop III dominated galaxies, and in their number counts.  This result is consistent with our previous work and reinforces our conclusion that this is best epoch to search for Pop III stars.
  
While observational constraints on the high redshift LF are uncertain \citep{2016PASA...33...37F, 2015MNRAS.450.3032M, 2015ApJ...803...34B, 2015ApJ...808..104O}, we find that our models predictions approximately follow the predicted faint end slope at $8 \le z \le 10$ and are in reasonable agreement with extrapolated Schechter functions to $z=12$.  We suspect that our simulations' resolution and relatively small volume limits our ability to accurately model galaxy counts at $z=15$ as they underperform both extrapolated Schechter models and \cite{2015ApJ...807L..12O} but are in reasonable agreement with LFs based on analytic models of star formation in high-redshift DM halos by \cite{0004-637X-813-1-21}. None of our simulated galaxies are brighter than $m_{\rm UV} = 31.4$ mag at $z>13$.

We note the largest differences in Pop III-bright galaxies, with at least 75\% of their flux coming from Pop III stars, occurs between the fiducial and P3SN simulations. The feedback effects of the Pop III log normal IMF coupled with the small number of Pop III stars that survive beyond 10 Myr result in far fewer galaxies with moderate Pop III flux fractions between 0.75 and $10^{-3}$.  The added feedback quenches subsequent star formation in median sized galaxies and below resulting in a largely bimodal `all-or-nothing' distribution of Pop III flux fractions for P3SN galaxies.  These galaxies vary between purely young, small Pop III galaxies and older, larger Pop II dominated galaxies that possess a tiny fraction of older, low-mass Pop III stars.

When considering the fraction of observable Pop III-bright galaxies, most of the simulations produce statistics very similar to the fid run. This run predicts no Pop III-bright galaxies across the redshift range studied. However the P3SN model stands out. While the fraction of Pop III-bright galaxies with $m_{\rm UV} \le 31.4$ mag is less than 1\% post-reionization, the fraction increases to approximately 19\% immediately before reionization. Additionally, the number of lensing opportunities for galaxies with $m_{\rm UV} \le 33$ mag for the P3SN run is encouraging at $z>9$ where we predict their occurrence at more than 1 in 10, although with weak statistics at $z>12$.

Turning to the chemical composition of our stars, we demonstrated that our P3SN simulation, with a log normal IMF for metal free stars with a characteristic mass of $60 \msun$, results in subsequent generations of stars with elemental abundances that reproduces the two populations of CEMP-no stars observed in the MW halo. In particular, our model reproduces the bimodal distribution of these stars where we see one population of CEMP-no stars likely polluted solely by Pop III SN with [C/H] $\approx -1$ and a second population likely polluted by both Pop III and Pop II SN with a higher overall metallicity and [C/Fe] $ \approx +0.7$. This supports the notion of a Pop III IMF characterized by stars in the $20 \msun \leq M_{\star} < 140 \msun$ range.  

We also integrated SN yields across subranges of our log normal IMF. We found that stars in the 20 to 120 $\msun$ mass range result in a subsequent generation of polluted stars that correlate well with observations of CEMP-no MW halo stars. Once again this supports the assumption that early minihalos generated a few Pop III stars with masses predominately in this range but possibly biased even more toward the heavier end of this range than predicted by a log normal distribution. Of course, there are still large uncertainties in the nucleosynthetic products of Pop III SN.

Including stars more massive than 140 $\msun$ in our IMF yields resulted in [C/Fe] $ \approx 0$ and a subsequent generation of Pop II stars that did not match observations. This result is at least in part due to the simulations' SP mass resolution. While as few as 1-4 Pop III stars may have been created in early minihalos, our simulations' SP mass resolution of $8.6\times10^3\, \msun$ means that stars greater than 140 $\msun$ were represented in all star forming galaxies. Increasing the mass resolution of our simulation would address this shortcoming.

A natural follow-on to this study would be to examine the effects of stochastically sampling the IMF across the range of masses to determine the effects of the various SN yields on the chemical composition of subsequent stellar generations. Such a study could help to further pin-down the Pop III IMF with regard to observations of metal poor stars. This, and subsequent studies, will help to further our understanding of the Pop III IMF and the physical processes relevant to modeling the evolution of high-redshift galaxies. 

\acknowledgments
We would like to thank Jinmi Yoon for supplying us with data for CEMP-no stars along with Rogier Windhorst and Seth Cohen for fruitful discussions and suggestions. Special thanks to, Alexander Heger for answering several questions about Pop III SN yields. 

This work was supported in by the National Science Foundation under Grants AST-1715876 \& PHY-1430152 (the Joint Institute for Nuclear Astrophysics -- Center for the Evolution of the Elements), and NASA theory grant NNX15AK82G. Benoit C\^ot\'e was also supported by the ERC Consolidator Grant (Hungary) funding scheme (project RADIOSTAR, GA n. 724560). The simulations and much of the analysis for this work was carried out using the NASA High-End Computing Capability (HECC) and the Texas Advanced Computing Center (TACC) at the University of Texas at Austin under grant TG-AST130021. We would also like to thank the NASA HECC support team.

\software{\textsc{ramses} \citep{2010ascl.soft11007T}, AdaptaHop \citep{2013ascl.soft05004C}, MUSIC \citep{2013ascl.soft11011H}, pynbody \citep{2013ascl.soft05002P}, SYGMA \citep{2018ascl.soft06019R, 2018ApJS..237...42R}, yt \citep{2010ascl.soft11022T}}

\clearpage
\begin{appendices}
\appendix{}\label{sec:append}

In this appendix we briefly compare the results from fid\_orig \citepalias{2018ApJ...854...75S}, fid, and, for Pop III-bright galaxies, the P3SN simulations. The results for fid and fid\_orig are based on the same \textsc{ramses} simulation data and hence used the same SPs, at each redshift. Specifically, the surviving fraction of SP mass after 10 Myr, 90\%, corresponds to a Salpeter IMF. The analysis of the fid simulation data also assumed this distribution of mass, in post-processing, for the surviving stars when computing flux. Hence most of the stellar mass was concentrated in smaller, long-lived stars. However, for fid\_orig, we assumed a log normal distribution of stellar mass for the surviving fraction of each SP resulting in a much larger fraction of massive, short-lived stars. Run P3SN modeled a log normal IMF within \textsc{ramses} and hence 99\% of each SP's mass was lost to SNe after 10 Myr. Even with this reduction in mass, the fraction of massive young stars results in flux comparable to the fid case as depicted in Figure \ref{fig:lfFidP3sn}.

\begin{figure}[b!]
\begin{center}
\includegraphics[width=0.8\textwidth]{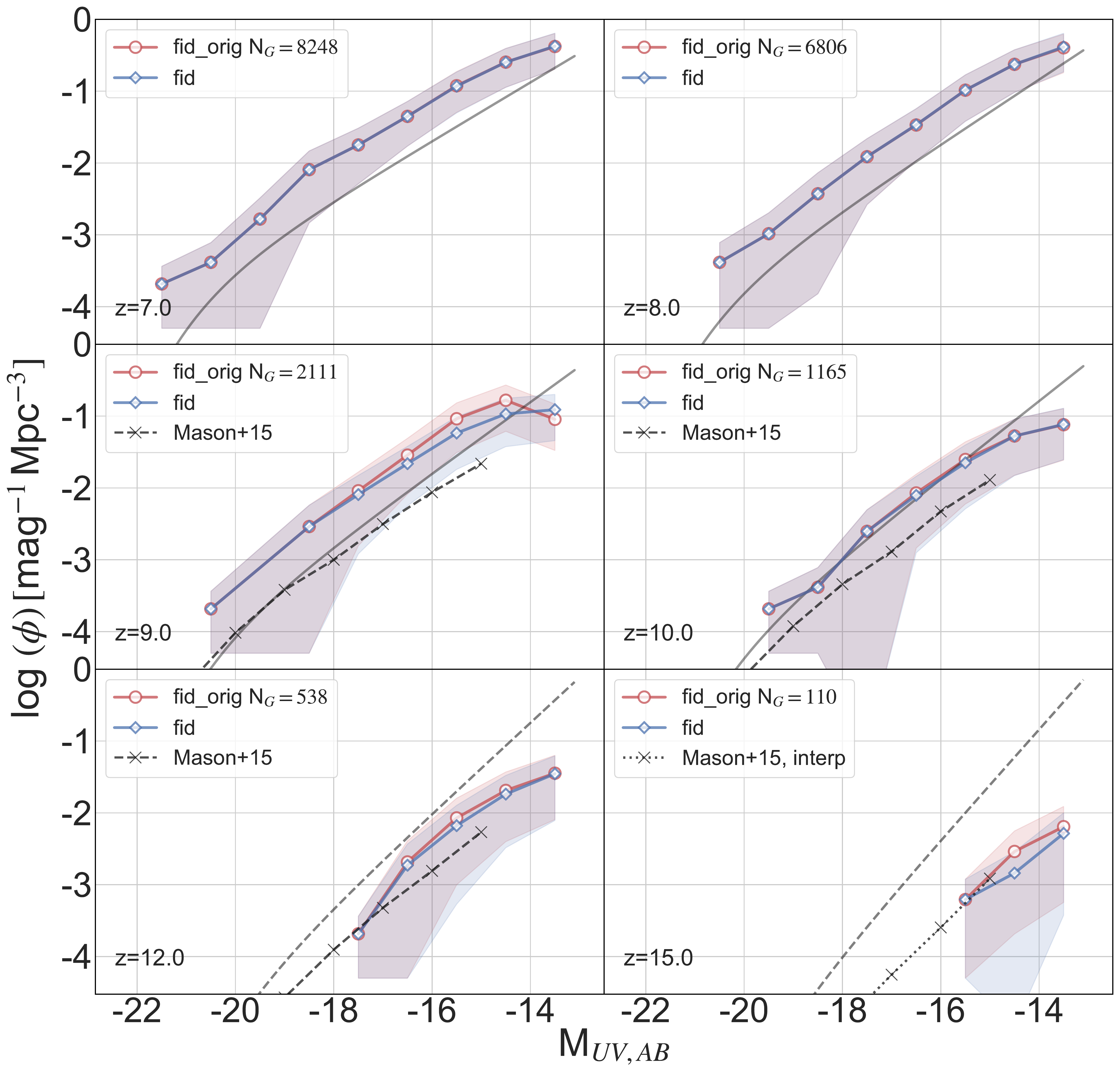} 
\caption{UV LFs derived from our simulations with 1$\sigma$ error bounds including both Poisson noise and sample variance. Solid grey lines are \cite{2016PASA...33...37F} Schechter fits. Dashed grey are extrapolated. We also include LFs from \cite{0004-637X-813-1-21} for comparison. fid\_orig uses the log normal Pop III IMF SEDs from \cite{2018ApJ...854...75S}, while fid uses the Salpeter IMF SEDs. Although the interpretation of surviving SP mass is very different in these two models, the resulting overall LF are nearly indistinguishable.  }
\end{center}
\label{fig:lfOrigNew}
\end{figure}

Figure \ref{fig:lfOrigNew}, compares the LFs for fid and fid\_orig. As depicted, the change in the IMF used to interpret stellar mass -- along with the corresponding SEDs for Pop III stars  -- results in only very small changes to the LFs of our galaxies. It appears that the increased luminosity of the small fraction of massive, young Pop III stars in the fid\_orig log normal IMF is mostly matched by the extra luminosity of the larger fraction of older Pop III stars in the fid Salpeter IMF.

\begin{figure}[h!]
\begin{center}
\includegraphics[width=0.450\textwidth]{./PopIII_fraction-HistFid_Sal-trim.pdf} 
\includegraphics[width=0.50\textwidth]{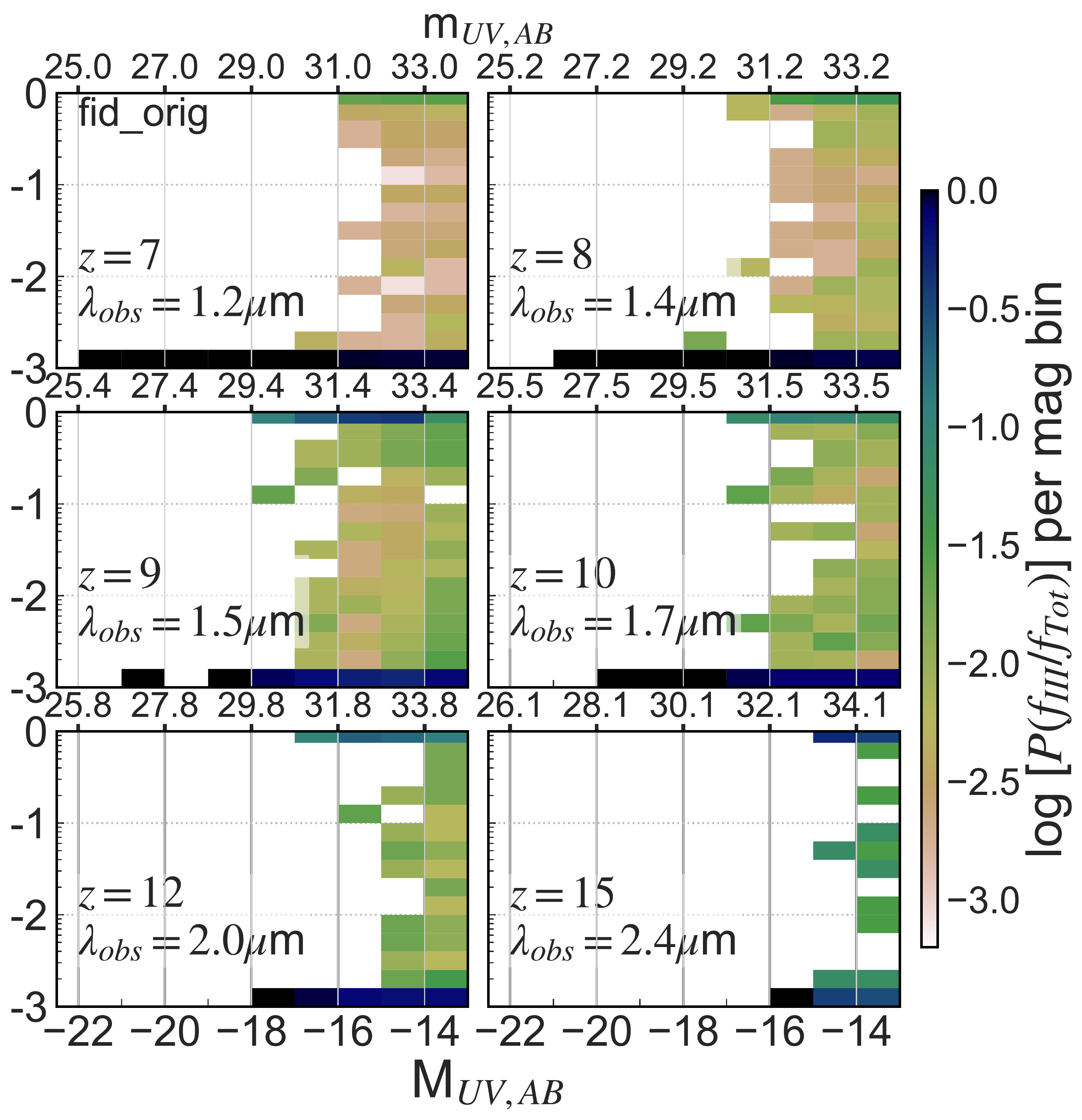}
\caption{Same as Figure \ref{fig:p3Prob2} but comparing fid (\textit{left}) to fid\_orig (\textit{right}). The fid run modeled Pop III flux derived from a Salpeter IMF. The fid\_orig simulation used a log normal IMF \citepalias{2018ApJ...854...75S} for Pop III stars. SEDs based on a Salpeter IMF for Pop III stars result in more flux from smaller, older stars and result in a large fraction of hybrid galaxies ($10^{-3} < \nicefrac{f_{\rm III}}{f_{\rm Tot}} < 0.75$). The log normal case, fid\_orig, tends to be more bimodal, especially at $z\ge10$ where ongoing Pop III star formation results in a larger fraction of young, massive stars.  $\lambda_{obs}$ indicates the observational wavelength of the 1500 $\AA$ reference. \label{fig:p3Prob1}}
\end{center}
\end{figure}

Figure \ref{fig:p3Prob1} depicts the Pop III flux fraction as a function of the magnitude and redshift of our galaxies for fid (left) and fid\_orig (right).  We have normalized the probabilities in each magnitude bin such that they sum to one. Pop III flux from hybrid galaxies, with $10^{-3} < \nicefrac{f_{\rm III}}{f_{\rm Tot}} < 0.5$, in the fid run is more prominent than in the fid\_orig run since the Salpeter IMF results in a larger fraction of Pop III stellar mass mapped to smaller, longer lived stars. Such stars result in moderate fractions of Pop III flux.  Conversely, galaxies with $\nicefrac{f_{\rm III}}{f_{\rm Tot}} \ge 0.75$ are less populous in this model since less mass was mapped to younger, massive stars. So while the IMF used in conjunction with the SEDs does not significantly effect the luminosity of our galaxies, it does make a dramatic different in the type of flux they generate.
\end{appendices}

\begin{figure*}[h]
\begin{centering}
\includegraphics[width=.975\textwidth]{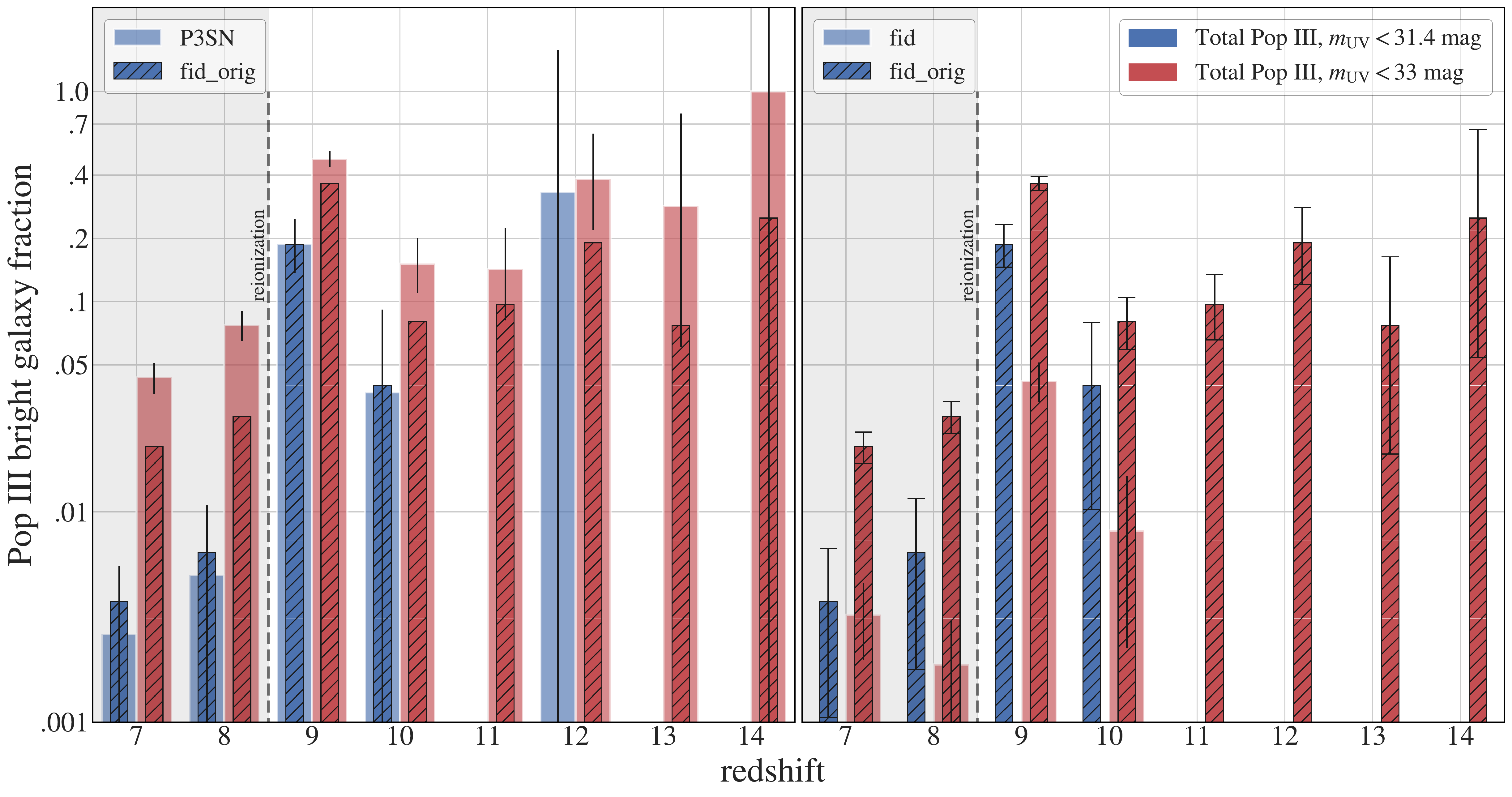} 
\caption{A comparison of the fraction of Pop III-bright galaxies for the P3SN and fid runs to the fid\_orig run, with Poisson error bars. Each panel indicates the joint probability of finding an observable, $m_{\rm UV} \le 31.4$ mag (blue) or $m_{\rm UV} \le 33$ mag (red),  Pop III-bright galaxy as a fraction of all galaxies meeting the magnitude cutoff.  Run P3SN closely matches the fid\_orig case since both of the runs model flux as coming from a log normal distribution of stars for each SP. While fid and fid\_orig have the same SP mass at each redshift, they each modeled the flux coming from the constituent stars using a different IMF: Salpeter for fid and log normal for fid\_orig. The flux from massive stars, as modeled in fid\_orig, results in a much larger fraction of Pop III-bright galaxies than from the stars following a Salpeter IMF in fid.\label{fig:p3brightOrig}}
\end{centering}
\end{figure*}

Figure \ref{fig:p3brightOrig} compares the fraction of Pop III-bright galaxies for the P3SN and fid runs to fid\_orig. We note that the fid\_orig\footnote{Fig. \ref{fig:p3brightOrig} presents the corrected version of fid\_orig data that was plotted incorrectly in \citetalias{2018ApJ...854...75S}. The errors in the plot, and incorrect percentages mentioned in that discussion, do not effect any of our conclusions in that work.} run generated more, or as many, Pop III-bright galaxies at $m_{\rm UV} \le 31.4$ mag over the range $7 \le z \le 10$ than P3SN. The smaller SN fraction, 10\%, in the fid\_orig resulted in more mass in stars than in the P3SN run. The fid case generated no Pop III-bright galaxies over this range. Additionally, the fraction of Pop III-bright galaxies with $m_{\rm UV} \le 33$ mag in the fid run drops by an average factor of 10 as compared to fid\_orig over the same redshift range due to the reduced fraction of mass in massive, luminous stars.

\clearpage
\bibliographystyle{apj} 
\bibliography{Paper3}

\begin{thebibliography}{}
\expandafter\ifx\csname natexlab\endcsname\relax\def\natexlab#1{#1}\fi

\bibitem[{Abel {et~al.}(2000)Abel, Bryan, \& Norman}]{abel2000formation}
Abel, T., Bryan, G.~L., \& Norman, M.~L. 2000, \apj, 540, 39

\bibitem[{Abel {et~al.}(2002)Abel, Bryan, \& Norman}]{abel2002formation}
---. 2002, science, 295, 93

\bibitem[{{Aubert} {et~al.}(2004){Aubert}, {Pichon}, \&
  {Colombi}}]{2004MNRAS.352..376A}
{Aubert}, D., {Pichon}, C., \& {Colombi}, S. 2004, \mnras, 352, 376

\bibitem[{{Barai} \& {de Gouveia Dal Pino}(2018)}]{2018arXiv180704768B}
{Barai}, P., \& {de Gouveia Dal Pino}, E.~M. 2018, ArXiv e-prints,
  arXiv:1807.04768

\bibitem[{{Beers} \& {Christlieb}(2005)}]{2005ARA&A..43..531B}
{Beers}, T.~C., \& {Christlieb}, N. 2005, \araa, 43, 531

\bibitem[{{Bose} {et~al.}(2018){Bose}, {Deason}, \&
  {Frenk}}]{2018arXiv180210096B}
{Bose}, S., {Deason}, A.~J., \& {Frenk}, C.~S. 2018, ArXiv e-prints,
  arXiv:1802.10096

\bibitem[{{Bouwens} {et~al.}(2015){Bouwens}, {Illingworth}, {Oesch}, {Trenti},
  {Labb{\'e}}, {Bradley}, {Carollo}, {van Dokkum}, {Gonzalez}, {Holwerda},
  {Franx}, {Spitler}, {Smit}, \& {Magee}}]{2015ApJ...803...34B}
{Bouwens}, R.~J., {Illingworth}, G.~D., {Oesch}, P.~A., {et~al.} 2015, \apj,
  803, 34

\bibitem[{{Bromm} {et~al.}(1999){Bromm}, {Coppi}, \&
  {Larson}}]{1999ApJ...527L...5B}
{Bromm}, V., {Coppi}, P.~S., \& {Larson}, R.~B. 1999, \apjl, 527, L5

\bibitem[{Bromm {et~al.}(2002)Bromm, Coppi, \& Larson}]{Bromm2002}
Bromm, V., Coppi, P.~S., \& Larson, R.~B. 2002, \apj, 564, 23

\bibitem[{{Bromm} {et~al.}(2001){Bromm}, {Ferrara}, {Coppi}, \&
  {Larson}}]{2001MNRAS.328..969B}
{Bromm}, V., {Ferrara}, A., {Coppi}, P.~S., \& {Larson}, R.~B. 2001, \mnras,
  328, 969

\bibitem[{{Brook} {et~al.}(2007){Brook}, {Kawata}, {Scannapieco}, {Martel}, \&
  {Gibson}}]{2007ApJ...661...10B}
{Brook}, C.~B., {Kawata}, D., {Scannapieco}, E., {Martel}, H., \& {Gibson},
  B.~K. 2007, \apj, 661, 10

\bibitem[{{Cassata} {et~al.}(2013){Cassata}, {Le F{\`e}vre}, {Charlot},
  {Contini}, {Cucciati}, {Garilli}, {Zamorani}, {Adami}, {Bardelli}, {Le Brun},
  {Lemaux}, {Maccagni}, {Pollo}, {Pozzetti}, {Tresse}, {Vergani}, {Zanichelli},
  \& {Zucca}}]{2013A&A...556A..68C}
{Cassata}, P., {Le F{\`e}vre}, O., {Charlot}, S., {et~al.} 2013, \aap, 556, A68

\bibitem[{{Castellani} {et~al.}(1983){Castellani}, {Chieffi}, \&
  {Tornambe}}]{1983ApJ...272..249C}
{Castellani}, V., {Chieffi}, A., \& {Tornambe}, A. 1983, \apj, 272, 249

\bibitem[{{Clark} {et~al.}(2011){Clark}, {Glover}, {Klessen}, \&
  {Bromm}}]{2011ApJ...727..110C}
{Clark}, P.~C., {Glover}, S.~C.~O., {Klessen}, R.~S., \& {Bromm}, V. 2011,
  \apj, 727, 110

\bibitem[{{Colombi}(2013)}]{2013ascl.soft05004C}
{Colombi}, S. 2013, {AdaptaHOP: Subclump finder}, Astrophysics Source Code
  Library, ascl:1305.004

\bibitem[{{Couchman} \& {Rees}(1986)}]{1986MNRAS.221...53C}
{Couchman}, H.~M.~P., \& {Rees}, M.~J. 1986, \mnras, 221, 53

\bibitem[{{de Bennassuti} {et~al.}(2017){de Bennassuti}, {Salvadori},
  {Schneider}, {Valiante}, \& {Omukai}}]{2017MNRAS.465..926D}
{de Bennassuti}, M., {Salvadori}, S., {Schneider}, R., {Valiante}, R., \&
  {Omukai}, K. 2017, \mnras, 465, 926

\bibitem[{Deharveng {et~al.}(2010)Deharveng, Schuller, Anderson, Zavagno,
  Wyrowski, Menten, Bronfman, Testi, Walmsley, \&
  Wienen}]{deharveng2010gallery}
Deharveng, L., Schuller, F., Anderson, L., {et~al.} 2010, \aap, 523, A6

\bibitem[{{Dopcke} {et~al.}(2011){Dopcke}, {Glover}, {Clark}, \&
  {Klessen}}]{2011ApJ...729L...3D}
{Dopcke}, G., {Glover}, S.~C.~O., {Clark}, P.~C., \& {Klessen}, R.~S. 2011,
  \apjl, 729, L3

\bibitem[{{Dopcke} {et~al.}(2013){Dopcke}, {Glover}, {Clark}, \&
  {Klessen}}]{2013ApJ...766..103D}
---. 2013, \apj, 766, 103

\bibitem[{{Dubois} \& {Teyssier}(2008)}]{2008A&A...477...79D}
{Dubois}, Y., \& {Teyssier}, R. 2008, \aap, 477, 79

\bibitem[{{El Eid} {et~al.}(1983){El Eid}, {Fricke}, \&
  {Ober}}]{1983A&A...119...54E}
{El Eid}, M.~F., {Fricke}, K.~J., \& {Ober}, W.~W. 1983, \aap, 119, 54

\bibitem[{{Ferland} {et~al.}(1999){Ferland}, {van Hoof}, {Verner}, {Verner},
  {Ferguson}, {Hamann}, {Kingdon}, {Korista}, \&
  {Shields}}]{1999ascl.soft10001F}
{Ferland}, G., {van Hoof}, P., {Verner}, D., {et~al.} 1999, {Cloudy: Numerical
  simulation of plasmas and their spectra}, Astrophysics Source Code Library,
  ascl:9910.001

\bibitem[{{Finkelstein}(2016)}]{2016PASA...33...37F}
{Finkelstein}, S.~L. 2016, \pasa, 33, e037

\bibitem[{{Frebel} \& {Norris}(2015)}]{2015ARA&A..53..631F}
{Frebel}, A., \& {Norris}, J.~E. 2015, \araa, 53, 631

\bibitem[{{Frebel} {et~al.}(2005){Frebel}, {Aoki}, {Christlieb}, {Ando},
  {Asplund}, {Barklem}, {Beers}, {Eriksson}, {Fechner}, {Fujimoto}, {Honda},
  {Kajino}, {Minezaki}, {Nomoto}, {Norris}, {Ryan}, {Takada-Hidai},
  {Tsangarides}, \& {Yoshii}}]{2005Natur.434..871F}
{Frebel}, A., {Aoki}, W., {Christlieb}, N., {et~al.} 2005, \nat, 434, 871

\bibitem[{{Freeman} \& {Bland-Hawthorn}(2002)}]{2002ARA&A..40..487F}
{Freeman}, K., \& {Bland-Hawthorn}, J. 2002, \araa, 40, 487

\bibitem[{{Furlanetto} \& {Oh}(2008)}]{2008ApJ...682...14F}
{Furlanetto}, S.~R., \& {Oh}, S.~P. 2008, \apj, 682, 14

\bibitem[{{Gardner} {et~al.}(2006){Gardner}, {Mather}, {Clampin}, {Doyon},
  {Greenhouse}, {Hammel}, {Hutchings}, {Jakobsen}, {Lilly}, {Long}, {Lunine},
  {McCaughrean}, {Mountain}, {Nella}, {Rieke}, {Rieke}, {Rix}, {Smith},
  {Sonneborn}, {Stiavelli}, {Stockman}, {Windhorst}, \&
  {Wright}}]{2006SSRv..123..485G}
{Gardner}, J.~P., {Mather}, J.~C., {Clampin}, M., {et~al.} 2006, \ssr, 123, 485

\bibitem[{{Haardt} \& {Madau}(1996)}]{1996ApJ...461...20H}
{Haardt}, F., \& {Madau}, P. 1996, \apj, 461, 20

\bibitem[{{Hahn} \& {Abel}(2013)}]{2013ascl.soft11011H}
{Hahn}, O., \& {Abel}, T. 2013, {MUSIC: MUlti-Scale Initial Conditions},
  Astrophysics Source Code Library, ascl:1311.011

\bibitem[{Hartwig {et~al.}(2014)Hartwig, Bromm, Klessen, \&
  Glover}]{Hartwig2014}
Hartwig, T., Bromm, V., Klessen, R.~S., \& Glover, S. C.~O. 2014, \mnras, 447,
  3892

\bibitem[{{Hartwig} {et~al.}(2015){Hartwig}, {Bromm}, {Klessen}, \&
  {Glover}}]{2015MNRAS.447.3892H}
{Hartwig}, T., {Bromm}, V., {Klessen}, R.~S., \& {Glover}, S.~C.~O. 2015,
  \mnras, 447, 3892

\bibitem[{{Hartwig} {et~al.}(2018){Hartwig}, {Yoshida}, {Magg}, {Frebel},
  {Glover}, {G{\'o}mez}, {Griffen}, {Ishigaki}, {Ji}, {Klessen}, {O'Shea}, \&
  {Tominaga}}]{2018MNRAS.478.1795H}
{Hartwig}, T., {Yoshida}, N., {Magg}, M., {et~al.} 2018, \mnras, 478, 1795

\bibitem[{{Heger}(2018)}]{hegerwebsite}
{Heger}, A. 2018, {STARFIT}, \url{http://starfit.org/}, accessed: 2018-09-12

\bibitem[{{Heger} {et~al.}(2003){Heger}, {Fryer}, {Woosley}, {Langer}, \&
  {Hartmann}}]{2003ApJ...591..288H}
{Heger}, A., {Fryer}, C.~L., {Woosley}, S.~E., {Langer}, N., \& {Hartmann},
  D.~H. 2003, \apj, 591, 288

\bibitem[{{Heger} \& {Woosley}(2002)}]{2002ApJ...567..532H}
{Heger}, A., \& {Woosley}, S.~E. 2002, \apj, 567, 532

\bibitem[{{Heger} \& {Woosley}(2010)}]{2010ApJ...724..341H}
---. 2010, \apj, 724, 341

\bibitem[{Hirano \& Yoshida(2013)}]{hirano2013radiative}
Hirano, S., \& Yoshida, N. 2013, \apj, 763, 52

\bibitem[{{Ishigaki} {et~al.}(2018){Ishigaki}, {Tominaga}, {Kobayashi}, \&
  {Nomoto}}]{2018ApJ...857...46I}
{Ishigaki}, M.~N., {Tominaga}, N., {Kobayashi}, C., \& {Nomoto}, K. 2018, \apj,
  857, 46

\bibitem[{Jaacks {et~al.}(2018)Jaacks, Thompson, Finkelstein, \&
  Bromm}]{Jaacks2018}
Jaacks, J., Thompson, R., Finkelstein, S.~L., \& Bromm, V. 2018, \mnras, 475,
  4396

\bibitem[{{Jeon} {et~al.}(2012){Jeon}, {Pawlik}, {Greif}, {Glover}, {Bromm},
  {Milosavljevi{\'c}}, \& {Klessen}}]{2012AIPC.1480..325J}
{Jeon}, M., {Pawlik}, A.~H., {Greif}, T.~H., {et~al.} 2012, in American
  Institute of Physics Conference Series, Vol. 1480, American Institute of
  Physics Conference Series, ed. M.~{Umemura} \& K.~{Omukai}, 325--328

\bibitem[{{Johnson}(2010)}]{2010MNRAS.404.1425J}
{Johnson}, J.~L. 2010, \mnras, 404, 1425

\bibitem[{Johnson \& Bromm(2006)}]{johnson2006cooling}
Johnson, J.~L., \& Bromm, V. 2006, \mnras, 366, 247

\bibitem[{{Johnson} {et~al.}(2013){Johnson}, {Dalla Vecchia}, \&
  {Khochfar}}]{2013MNRAS.428.1857J}
{Johnson}, J.~L., {Dalla Vecchia}, C., \& {Khochfar}, S. 2013, \mnras, 428,
  1857

\bibitem[{Johnson {et~al.}(2013)Johnson, Vecchia, \& Khochfar}]{Johnson2013a}
Johnson, J.~L., Vecchia, C.~D., \& Khochfar, S. 2013, \mnras, 428, 1857

\bibitem[{{Kashikawa} {et~al.}(2012){Kashikawa}, {Nagao}, {Toshikawa},
  {Ishizaki}, {Egami}, {Hayashi}, {Ly}, {Malkan}, {Matsuda}, {Shimasaku},
  {Iye}, {Ota}, {Shibuya}, {Jiang}, {Taniguchi}, \&
  {Shioya}}]{2012ApJ...761...85K}
{Kashikawa}, N., {Nagao}, T., {Toshikawa}, J., {et~al.} 2012, \apj, 761, 85

\bibitem[{{Keller} {et~al.}(2014){Keller}, {Bessell}, {Frebel}, {Casey},
  {Asplund}, {Jacobson}, {Lind}, {Norris}, {Yong}, {Heger}, {Magic}, {da
  Costa}, {Schmidt}, \& {Tisserand}}]{2014Natur.506..463K}
{Keller}, S.~C., {Bessell}, M.~S., {Frebel}, A., {et~al.} 2014, \nat, 506, 463

\bibitem[{{Kitayama} \& {Yoshida}(2005)}]{2005ApJ...630..675K}
{Kitayama}, T., \& {Yoshida}, N. 2005, \apj, 630, 675

\bibitem[{Komatsu {et~al.}(2009)Komatsu, Dunkley, Nolta, Bennett, Gold,
  Hinshaw, Jarosik, Larson, Limon, Page, {et~al.}}]{2011ApJS..192...18K}
Komatsu, E., Dunkley, J., Nolta, M., {et~al.} 2009, \apjs, 180, 330

\bibitem[{Larson(1973)}]{larson1973simple}
Larson, R.~B. 1973, \mnras, 161, 133

\bibitem[{{Leitherer} {et~al.}(2014){Leitherer}, {Ekstr{\"o}m}, {Meynet},
  {Schaerer}, {Agienko}, \& {Levesque}}]{2014ApJS..212...14L}
{Leitherer}, C., {Ekstr{\"o}m}, S., {Meynet}, G., {et~al.} 2014, \apjs, 212, 14

\bibitem[{{Loeb}(2010)}]{2010hdfs.book.....L}
{Loeb}, A. 2010, {How Did the First Stars and Galaxies Form?}

\bibitem[{{Mackey} {et~al.}(2003){Mackey}, {Bromm}, \&
  {Hernquist}}]{2003ApJ...586....1M}
{Mackey}, J., {Bromm}, V., \& {Hernquist}, L. 2003, \apj, 586, 1

\bibitem[{{Madau}(1995)}]{1995ApJ...441...18M}
{Madau}, P. 1995, \apj, 441, 18

\bibitem[{{Madau} \& {Dickinson}(2014)}]{2014ARA&A..52..415M}
{Madau}, P., \& {Dickinson}, M. 2014, \araa, 52, 415

\bibitem[{Martin {et~al.}(1996)Martin, Schwarz, \& Mandy}]{martin1996master}
Martin, P., Schwarz, D., \& Mandy, M. 1996, \apj, 461, 265

\bibitem[{Mason {et~al.}(2015)Mason, Trenti, \& Treu}]{0004-637X-813-1-21}
Mason, C.~A., Trenti, M., \& Treu, T. 2015, The Astrophysical Journal, 813, 21

\bibitem[{{McLeod} {et~al.}(2015){McLeod}, {McLure}, {Dunlop}, {Robertson},
  {Ellis}, \& {Targett}}]{2015MNRAS.450.3032M}
{McLeod}, D.~J., {McLure}, R.~J., {Dunlop}, J.~S., {et~al.} 2015, \mnras, 450,
  3032

\bibitem[{{Nagao} {et~al.}(2008){Nagao}, {Sasaki}, {Maiolino}, {Grady},
  {Kashikawa}, {Ly}, {Malkan}, {Motohara}, {Murayama}, {Schaerer}, {Shioya}, \&
  {Taniguchi}}]{2008ApJ...680..100N}
{Nagao}, T., {Sasaki}, S.~S., {Maiolino}, R., {et~al.} 2008, \apj, 680, 100

\bibitem[{{Nomoto} {et~al.}(2013){Nomoto}, {Kobayashi}, \&
  {Tominaga}}]{2013ARA&A..51..457N}
{Nomoto}, K., {Kobayashi}, C., \& {Tominaga}, N. 2013, \araa, 51, 457

\bibitem[{{Norman}(2010)}]{2010AIPC.1294...17N}
{Norman}, M.~L. 2010, in American Institute of Physics Conference Series, Vol.
  1294, American Institute of Physics Conference Series, ed. D.~J. {Whalen},
  V.~{Bromm}, \& N.~{Yoshida}, 17--27

\bibitem[{{Oesch} {et~al.}(2015){Oesch}, {Bouwens}, {Illingworth}, {Franx},
  {Ammons}, {van Dokkum}, {Trenti}, \& {Labb{\'e}}}]{2015ApJ...808..104O}
{Oesch}, P.~A., {Bouwens}, R.~J., {Illingworth}, G.~D., {et~al.} 2015, \apj,
  808, 104

\bibitem[{{Oesch} {et~al.}(2013){Oesch}, {Bouwens}, {Illingworth}, {Labb{\'e}},
  {Franx}, {van Dokkum}, {Trenti}, {Stiavelli}, {Gonzalez}, \&
  {Magee}}]{2013ApJ...773...75O}
---. 2013, \apj, 773, 75

\bibitem[{{Omukai} {et~al.}(2005){Omukai}, {Tsuribe}, {Schneider}, \&
  {Ferrara}}]{2005ApJ...626..627O}
{Omukai}, K., {Tsuribe}, T., {Schneider}, R., \& {Ferrara}, A. 2005, \apj, 626,
  627

\bibitem[{O'Shea \& Norman(2007)}]{o2007population}
O'Shea, B.~W., \& Norman, M.~L. 2007, The Astrophysical Journal, 654, 66

\bibitem[{{O'Shea} {et~al.}(2015){O'Shea}, {Wise}, {Xu}, \&
  {Norman}}]{2015ApJ...807L..12O}
{O'Shea}, B.~W., {Wise}, J.~H., {Xu}, H., \& {Norman}, M.~L. 2015, \apjl, 807,
  L12

\bibitem[{{Pallottini} {et~al.}(2014){Pallottini}, {Ferrara}, {Gallerani},
  {Salvadori}, \& {D'Odorico}}]{2014MNRAS.440.2498P}
{Pallottini}, A., {Ferrara}, A., {Gallerani}, S., {Salvadori}, S., \&
  {D'Odorico}, V. 2014, \mnras, 440, 2498

\bibitem[{{Pan} \& {Scannapieco}(2010)}]{2010ApJ...721.1765P}
{Pan}, L., \& {Scannapieco}, E. 2010, \apj, 721, 1765

\bibitem[{{Pan} {et~al.}(2012){Pan}, {Scannapieco}, \&
  {Scalo}}]{2012JFM...700..459P}
{Pan}, L., {Scannapieco}, E., \& {Scalo}, J. 2012, Journal of Fluid Mechanics,
  700, 459

\bibitem[{{Pan} {et~al.}(2013){Pan}, {Scannapieco}, \&
  {Scalo}}]{2013ApJ...775..111P}
---. 2013, \apj, 775, 111

\bibitem[{{Panagia} {et~al.}(2003){Panagia}, {Stiavelli}, {Ferguson}, \&
  {Stockman}}]{2003RMxAC..17..230P}
{Panagia}, N., {Stiavelli}, M., {Ferguson}, H., \& {Stockman}, H.~S. 2003, in
  Revista Mexicana de Astronomia y Astrofisica, vol.~27, Vol.~17, Revista
  Mexicana de Astronomia y Astrofisica Conference Series, ed. V.~{Avila-Reese},
  C.~{Firmani}, C.~S. {Frenk}, \& C.~{Allen}, 230--234

\bibitem[{{Planck Collaboration} {et~al.}(2016){Planck Collaboration}, {Ade},
  {Aghanim}, {Arnaud}, {Ashdown}, {Aumont}, {Baccigalupi}, {Banday},
  {Barreiro}, {Bartlett}, \& et~al.}]{2016A&A...594A..13P}
{Planck Collaboration}, {Ade}, P.~A.~R., {Aghanim}, N., {et~al.} 2016, \aap,
  594, A13

\bibitem[{{Pontzen} {et~al.}(2013){Pontzen}, {Ro{\v s}kar}, {Stinson}, \&
  {Woods}}]{2013ascl.soft05002P}
{Pontzen}, A., {Ro{\v s}kar}, R., {Stinson}, G., \& {Woods}, R. 2013, {pynbody:
  N-Body/SPH analysis for python}, Astrophysics Source Code Library,
  ascl:1305.002

\bibitem[{Prieto {et~al.}(2017)Prieto, Escala, Volonteri, \&
  Dubois}]{2017ApJ...836..216P}
Prieto, J., Escala, A., Volonteri, M., \& Dubois, Y. 2017, \apj, 836, 216

\bibitem[{{Prieto} {et~al.}(2008){Prieto}, {Infante}, \&
  {Jimenez}}]{2008arXiv0809.2786P}
{Prieto}, J.~P., {Infante}, L., \& {Jimenez}, R. 2008, ArXiv e-prints,
  arXiv:0809.2786

\bibitem[{{Raiter} {et~al.}(2010){Raiter}, {Schaerer}, \&
  {Fosbury}}]{2010A&A...523A..64R}
{Raiter}, A., {Schaerer}, D., \& {Fosbury}, R.~A.~E. 2010, \aap, 523, A64

\bibitem[{Rasera \& Teyssier(2006)}]{2006A&A...445....1R}
Rasera, Y., \& Teyssier, R. 2006, \aap, 445, 1

\bibitem[{Reed {et~al.}(2005)Reed, Bower, Frenk, Gao, Jenkins, Theuns, \&
  White}]{2005MNRAS.363..393R}
Reed, D.~S., Bower, R., Frenk, C.~S., {et~al.} 2005, \mnras, 363, 393

\bibitem[{{Ritter} {et~al.}(2018{\natexlab{a}}){Ritter}, {C{\^o}t{\'e}},
  {Herwig}, {Navarro}, \& {Fryer}}]{2018ascl.soft06019R}
{Ritter}, C., {C{\^o}t{\'e}}, B., {Herwig}, F., {Navarro}, J.~F., \& {Fryer},
  C. 2018{\natexlab{a}}, {SYGMA: Modeling stellar yields for galactic
  modeling}, Astrophysics Source Code Library, ascl:1806.019

\bibitem[{{Ritter} {et~al.}(2018{\natexlab{b}}){Ritter}, {C{\^o}t{\'e}},
  {Herwig}, {Navarro}, \& {Fryer}}]{2018ApJS..237...42R}
{Ritter}, C., {C{\^o}t{\'e}}, B., {Herwig}, F., {Navarro}, J.~F., \& {Fryer},
  C.~L. 2018{\natexlab{b}}, \apjs, 237, 42

\bibitem[{Rosen \& Bregman(1995)}]{rosen1995global}
Rosen, A., \& Bregman, J.~N. 1995, \apj, 440, 634

\bibitem[{Salpeter(1955)}]{salpeter1955luminosity}
Salpeter, E.~E. 1955, \apj, 121, 161

\bibitem[{{Salvadori} {et~al.}(2010){Salvadori}, {Ferrara}, {Schneider},
  {Scannapieco}, \& {Kawata}}]{2010MNRAS.401L...5S}
{Salvadori}, S., {Ferrara}, A., {Schneider}, R., {Scannapieco}, E., \&
  {Kawata}, D. 2010, \mnras, 401, L5

\bibitem[{{Sarmento} {et~al.}(2018){Sarmento}, {Scannapieco}, \&
  {Cohen}}]{2018ApJ...854...75S}
{Sarmento}, R., {Scannapieco}, E., \& {Cohen}, S. 2018, \apj, 854, 75

\bibitem[{{Sarmento} {et~al.}(2017){Sarmento}, {Scannapieco}, \&
  {Pan}}]{2017ApJ...834...23S}
{Sarmento}, R., {Scannapieco}, E., \& {Pan}, L. 2017, \apj, 834, 23

\bibitem[{{Scannapieco} {et~al.}(2005){Scannapieco}, {Madau}, {Woosley},
  {Heger}, \& {Ferrara}}]{2005ApJ...633.1031S}
{Scannapieco}, E., {Madau}, P., {Woosley}, S., {Heger}, A., \& {Ferrara}, A.
  2005, \apj, 633, 1031

\bibitem[{{Scannapieco} \& {Oh}(2004)}]{2004ApJ...608...62S}
{Scannapieco}, E., \& {Oh}, S.~P. 2004, \apj, 608, 62

\bibitem[{{Scannapieco} {et~al.}(2003){Scannapieco}, {Schneider}, \&
  {Ferrara}}]{2003ApJ...589...35S}
{Scannapieco}, E., {Schneider}, R., \& {Ferrara}, A. 2003, \apj, 589, 35

\bibitem[{{Schaerer}(2002)}]{2002A&A...382...28S}
{Schaerer}, D. 2002, \aap, 382, 28

\bibitem[{{Schmidt}(1959)}]{1959ApJ...129..243S}
{Schmidt}, M. 1959, \apj, 129, 243

\bibitem[{{Schneider} {et~al.}(2012){Schneider}, {Omukai}, {Bianchi}, \&
  {Valiante}}]{2012MNRAS.419.1566S}
{Schneider}, R., {Omukai}, K., {Bianchi}, S., \& {Valiante}, R. 2012, \mnras,
  419, 1566

\bibitem[{{Simpson} {et~al.}(2013){Simpson}, {Bryan}, {Johnston}, {Smith}, {Mac
  Low}, {Sharma}, \& {Tumlinson}}]{2013MNRAS.432.1989S}
{Simpson}, C.~M., {Bryan}, G.~L., {Johnston}, K.~V., {et~al.} 2013, \mnras,
  432, 1989

\bibitem[{{Smidt} {et~al.}(2014){Smidt}, {Whalen}, {Wiggins}, {Even},
  {Johnson}, \& {Fryer}}]{2014ApJ...797...97S}
{Smidt}, J., {Whalen}, D.~J., {Wiggins}, B.~K., {et~al.} 2014, \apj, 797, 97

\bibitem[{{Sobral} {et~al.}(2015){Sobral}, {Matthee}, {Darvish}, {Schaerer},
  {Mobasher}, {R{\"o}ttgering}, {Santos}, \& {Hemmati}}]{2015ApJ...808..139S}
{Sobral}, D., {Matthee}, J., {Darvish}, B., {et~al.} 2015, \apj, 808, 139

\bibitem[{Somerville {et~al.}(2008)Somerville, Hopkins, Cox, Robertson, \&
  Hernquist}]{2008MNRAS.391..481S}
Somerville, R.~S., Hopkins, P.~F., Cox, T.~J., Robertson, B.~E., \& Hernquist,
  L. 2008, \mnras, 391, 481

\bibitem[{{Stacy} \& {Bromm}(2013)}]{2013MNRAS.433.1094S}
{Stacy}, A., \& {Bromm}, V. 2013, \mnras, 433, 1094

\bibitem[{Stacy \& Bromm(2014)}]{stacy2014first}
Stacy, A., \& Bromm, V. 2014, \apj, 785, 73

\bibitem[{{Stacy} {et~al.}(2010){Stacy}, {Greif}, \&
  {Bromm}}]{2010MNRAS.403...45S}
{Stacy}, A., {Greif}, T.~H., \& {Bromm}, V. 2010, \mnras, 403, 45

\bibitem[{{Susa}(2013)}]{2013ApJ...773..185S}
{Susa}, H. 2013, \apj, 773, 185

\bibitem[{{Susa} {et~al.}(2014){Susa}, {Hasegawa}, \&
  {Tominaga}}]{2014ApJ...792...32S}
{Susa}, H., {Hasegawa}, K., \& {Tominaga}, N. 2014, \apj, 792, 32

\bibitem[{{Teyssier}(2002)}]{2002A&A...385..337T}
{Teyssier}, R. 2002, \aap, 385, 337

\bibitem[{{Teyssier}(2010)}]{2010ascl.soft11007T}
---. 2010, {RAMSES: A new N-body and hydrodynamical code}, Astrophysics Source
  Code Library, ascl:1011.007

\bibitem[{{The yt project}(2010)}]{2010ascl.soft11022T}
{The yt project}. 2010, {yt: A Multi-Code Analysis Toolkit for Astrophysical
  Simulation Data}, Astrophysics Source Code Library, ascl:1011.022

\bibitem[{Timmes(2016)}]{Timmes}
Timmes, F.~X. 2016, The Composition of Galactic Gas at z=5, Private
  Communication

\bibitem[{{Tornatore} {et~al.}(2007){Tornatore}, {Ferrara}, \&
  {Schneider}}]{2007MNRAS.382..945T}
{Tornatore}, L., {Ferrara}, A., \& {Schneider}, R. 2007, \mnras, 382, 945

\bibitem[{{Tremblin} {et~al.}(2012){Tremblin}, {Audit}, {Minier}, {Schmidt}, \&
  {Schneider}}]{2012A&A...546A..33T}
{Tremblin}, P., {Audit}, E., {Minier}, V., {Schmidt}, W., \& {Schneider}, N.
  2012, \aap, 546, A33

\bibitem[{{Trenti} \& {Stiavelli}(2008)}]{2008ApJ...676..767T}
{Trenti}, M., \& {Stiavelli}, M. 2008, \apj, 676, 767

\bibitem[{{Tumlinson}(2006)}]{2006ApJ...641....1T}
{Tumlinson}, J. 2006, \apj, 641, 1

\bibitem[{{Tumlinson} {et~al.}(2001){Tumlinson}, {Giroux}, \&
  {Shull}}]{2001ApJ...550L...1T}
{Tumlinson}, J., {Giroux}, M.~L., \& {Shull}, J.~M. 2001, \apjl, 550, L1

\bibitem[{{Tumlinson} \& {Shull}(2000)}]{2000ApJ...528L..65T}
{Tumlinson}, J., \& {Shull}, J.~M. 2000, \apjl, 528, L65

\bibitem[{{Turk} {et~al.}(2009){Turk}, {Abel}, \&
  {O'Shea}}]{2009Sci...325..601T}
{Turk}, M.~J., {Abel}, T., \& {O'Shea}, B. 2009, Science, 325, 601

\bibitem[{Whalen {et~al.}(2004)Whalen, Abel, \& Norman}]{whalen2004radiation}
Whalen, D., Abel, T., \& Norman, M.~L. 2004, \apj, 610, 14

\bibitem[{{Whalen} {et~al.}(2008){Whalen}, {van Veelen}, {O'Shea}, \&
  {Norman}}]{2008ApJ...682...49W}
{Whalen}, D., {van Veelen}, B., {O'Shea}, B.~W., \& {Norman}, M.~L. 2008, \apj,
  682, 49

\bibitem[{Whalen(2012)}]{Whalen2012a}
Whalen, D.~J. 2012, 6

\bibitem[{{Whalen} {et~al.}(2013){Whalen}, {Joggerst}, {Fryer}, {Stiavelli},
  {Heger}, \& {Holz}}]{2013ApJ...768...95W}
{Whalen}, D.~J., {Joggerst}, C.~C., {Fryer}, C.~L., {et~al.} 2013, \apj, 768,
  95

\bibitem[{{Windhorst} {et~al.}(2018){Windhorst}, {Timmes}, {Wyithe},
  {Alpaslan}, {Andrews}, {Coe}, {Diego}, {Dijkstra}, {Driver}, {Kelly}, \&
  {Kim}}]{windhorst2018}
{Windhorst}, R.~A., {Timmes}, F.~X., {Wyithe}, J.~S.~B., {et~al.} 2018, \apjs,
  234, 41

\bibitem[{Wise {et~al.}(2011)Wise, Turk, Norman, \& Abel}]{wise2011birth}
Wise, J.~H., Turk, M.~J., Norman, M.~L., \& Abel, T. 2011, \apj, 745, 50

\bibitem[{{Woosley} {et~al.}(2002){Woosley}, {Heger}, \&
  {Weaver}}]{2002RvMP...74.1015W}
{Woosley}, S.~E., {Heger}, A., \& {Weaver}, T.~A. 2002, Reviews of Modern
  Physics, 74, 1015

\bibitem[{{Xu} {et~al.}(2016{\natexlab{a}}){Xu}, {Norman}, {O'Shea}, \&
  {Wise}}]{2016ApJ...823..140X}
{Xu}, H., {Norman}, M.~L., {O'Shea}, B.~W., \& {Wise}, J.~H.
  2016{\natexlab{a}}, \apj, 823, 140

\bibitem[{{Xu} {et~al.}(2016{\natexlab{b}}){Xu}, {Wise}, {Norman}, {Ahn}, \&
  {O'Shea}}]{2016ApJ...833...84X}
{Xu}, H., {Wise}, J.~H., {Norman}, M.~L., {Ahn}, K., \& {O'Shea}, B.~W.
  2016{\natexlab{b}}, \apj, 833, 84

\bibitem[{{Yoon} {et~al.}(2016){Yoon}, {Beers}, {Placco}, {Rasmussen},
  {Carollo}, {He}, {Hansen}, {Roederer}, \& {Zeanah}}]{2016ApJ...833...20Y}
{Yoon}, J., {Beers}, T.~C., {Placco}, V.~M., {et~al.} 2016, \apj, 833, 20

\bibitem[{{Yoon} {et~al.}(2018){Yoon}, {Beers}, {Dietz}, {Lee}, {Placco}, {Da
  Costa}, {Keller}, {Owen}, \& {Sharma}}]{2018arXiv180604738Y}
{Yoon}, J., {Beers}, T.~C., {Dietz}, S., {et~al.} 2018, ArXiv e-prints,
  arXiv:1806.04738

\end{thebibliography}

\end{document}